\documentclass[
   pre,aps,
   amsfonts,
   amssymb,
   amsmath,
   superscriptaddress,
   eqsecnum,
   showpacs,
   preprintnumbers,
   byrevtex]{revtex4-1}[12pt]

\usepackage{color}
\usepackage{morefloats}
\usepackage{graphicx}
\usepackage{subfigure}
\DeclareSymbolFont{AMSb}{U}{msb}{m}{n}
\DeclareSymbolFontAlphabet{\mathbb}{AMSb}

\usepackage{mathrsfs}
\usepackage{dcolumn}% Align table columns on decimal point
\begin{document}
\newcommand{\red}[1]{\textcolor{red}{#1}}  
\newcommand{\vphi}{\varphi}
\newcommand{\bq}{\begin{equation}}
\newcommand{\be}{\begin{equation}}
\newcommand{\ba}{\begin{eqnarray}}
\newcommand{\eq}{\end{equation}}
\newcommand{\ee}{\end{equation}}
\newcommand{\ea}{\end{eqnarray}}
\newcommand{\tchi} {{\tilde \chi}}
\newcommand{\tA} {{\tilde A}}
\newcommand{\sech} { {\rm sech}}
\newcommand{\pstar}{\mbox{$\psi^{\ast}$}}
\newcommand {\bPsi}{{\bar \Psi}}
\newcommand {\bpsi}{{\bar \psi}}
\newcommand {\barf}{{\bar f}}
\newcommand {\tu} {{\tilde u}}
\newcommand {\tv} {{\tilde v}}
\newcommand{\dq}{{\dot q}}
\newcommand {\tdelta} {{\tilde \delta}}
 \newcommand{\shao}[1]{\textcolor{red}{#1}}
 \newcommand{\etc}{\textit{etc}{}}
 \newcommand{\ie}{\textit{i.e.}{~}}
 \newcommand{\videpost}{\textit{vide post}{}}
 \newcommand{\eg}{\textit{e.g.}{~}}
 \newcommand{\ansatz}{\textit{ansatz}{ }}
 \newcommand{\tint}{\!\int\!}                   % "tight" int
\newcommand{\tintx}{\!\int\!\mathrm{d}x\,}     % "tight" int dd{x}
\newcommand{\tinty}{\!\int\!\mathrm{d}y\,}     % "tight" int dd{y}
\newcommand{\tintz}{\!\int\!\mathrm{d}z\,}     % "tight" int dd{z}
\newcommand{\rmi}{{\rm i}}  
\newcommand{\dd}[1]{\,\mathrm{d}#1\,}
\newcommand{\tpsi}{\tilde{\psi}}               % tilde psi 
\newcommand{\qc}{\>,\qquad}
\newcommand{\qcs}{\>,\quad}
\newcommand{\tphi}{\tilde{\phi}}   
\newcommand{\dv}[2]{\frac{\mathrm{d}#1}{\mathrm{d} #2}}
\newcommand{\pdv}[2]{\frac{\partial #1}{\partial #2}}
\newcommand{\PoissonB}[2]{\ensuremath{ \lbrace \,#1,#2\, \rbrace }}  % Poisson bracket
\maxdeadcycles=1000

%\preprint{LA-UR 19- (Draft parametricnlde.tex) : \today}
\newpage

%\tableofcontents{}

\newpage

\title{Parametrically driven nonlinear Dirac equation 
with arbitrary nonlinearity}

\author{Fred Cooper}
\email{cooper@santafe.edu}
\affiliation{Santa Fe Institute, Santa Fe, NM 87501, USA}
\affiliation{Theoretical Division and Center for Nonlinear Studies, 
Los Alamos National Laboratory, Los Alamos, New Mexico 87545, USA}

\author{Avinash Khare} 
\email{khare@physics.unipune.ac.in}
\affiliation{ Physics Department, 
   Savitribai Phule Pune University, 
   Pune 411007, India}
    \author{Niurka R.\ Quintero}
 \email{niurka@us.es}
 \affiliation{Departamento de F\'\i sica Aplicada I, E.P.S., Universidad de
 	Sevilla, Virgen de \'Africa 7, 41011, Sevilla, Spain}
  \author{Bernardo S\'anchez-Rey} 
  \email{bernardo@us.es}
  \affiliation{Departamento de F\'\i sica Aplicada I, E.P.S., Universidad de
 	Sevilla, Virgen de \'Africa 7, 41011, Sevilla, Spain}
\author{Franz G.  Mertens}
\email{franzgmertens@gmail.com}
\affiliation{Physikalisches Institut, Universit\"at Bayreuth, D-95440 Bayreuth, Germany} 

\author{Avadh Saxena}
\email{avadh@lanl.gov}
\affiliation{Theoretical Division and Center for Nonlinear Studies, 
Los Alamos National Laboratory, Los Alamos, New Mexico 87545, USA}
\vspace{10pt}
\date{\today}

\begin{abstract}
The damped and parametrically driven nonlinear Dirac equation with arbitrary nonlinearity parameter $\kappa$  is analyzed, when the external force is periodic in space  and given by $f(x) =r\cos(K x)$,  both numerically and in a variational approximation using five collective coordinates (time dependent shape parameters of the wave function). 
Our variational approximation  satisfies exactly the low-order moment equations.  Because of competition between the spatial period of the external force  $\lambda=2 \pi/K$, and the soliton width  $l_s$, which is a function of the nonlinearity $\kappa$ as well as the initial frequency $\omega_0$ of the solitary wave,  there is a transition (at fixed $\omega_0$)  from trapped to unbound behavior of the soliton, which depends on the parameters $r$ and $K$ of the external force and the nonlinearity parameter $\kappa$.  We previously studied this phenomena when $\kappa=1$ 
(2019 J. Phys. A: Math. Theor. {\bf 52}  285201)  where we showed that for $\lambda  \gg  l_s$ the soliton oscillates in an effective potential, while for $\lambda \ll l_s$  it moves uniformly as a free particle.  
In this paper we focus on the $\kappa$ dependence of the transition from oscillatory to particle behavior and explicitly compare the curves of the transition regime found in the collective coordinate approximation as a function of $r$ and $K$ when $\kappa=1/2,1,2$ at fixed value of the frequency $\omega_0$.  Since the solitary wave gets narrower for fixed $\omega_0$ as a function of $\kappa$, we expect and indeed find that the regime where the solitary wave is trapped is extended as we increase $\kappa$. 
\end{abstract}

\maketitle

%\tableofcontents

%\Large
%%%%%%%%%%%%%%%%%%%%%%%%%%%%%%%%%%%%%%%%%%%%%%%%%%%%%%%%%%%%%%%%%%%%%%
\section{Introduction}
%%%%%%%%%%%%%%%%%%%%%%%%%%%%%%%%%%%%%%%%%%%%%%%%%%%%%%%%%%%%%%%%%%%%%%

 The nonlinear Dirac (NLD) equation has had a long history in particle physics as a model field theory to describe the low energy behavior of the weak interactions starting with Fermi's theory of beta decay \cite{Fermi:1938}.  This theory  was recast by Feynman and Gell-Mann \cite{Feynman:1958}  as a nonlinear quantum field theory  with vector and axial vector 4-Fermi interactions. At the classical level, 
 the  Dirac equation was  generalized to include local self-interactions by Ivanenko  \cite{Ivanenko:1938}. Classical versions of  the nonlinear Dirac (NLD) equation with different self-interaction terms and in different spatial dimensions   have  found many applications as models  for various  physical systems such as a model for  extended particles   
\cite{Finkelstein:1951, finkelstein:1956,heisenberg:1957},  as a way of describing nonlinear optics \cite{barashenkov:1998}, 
 optical realizations of relativistic quantum mechanics 
\cite{longhi:2010,dreisow:2010,tran:2014},
in  honeycomb optical lattices hosting  Bose-Einstein condensates \cite{haddad:2009}, among others. 

% Different  NLD equations results from different self-interactions which preserve Lorentz invariance. These include
% the 
%scalar bilinear covariant \cite{gursey:1956,soler:1970,gross:1974,mathieu:1984},
%the vector bilinear covariant \cite{thirring:1958} 
%and the axial vector bilinear covariant \cite{mathieu:1983}.  Moreover, models involving 
%both scalar and pseudoscalar bilinear covariants \cite{ranada:1984} 
%as well as both scalar and vector bilinear covariants \cite{stubbe:1986,nogami:1992} have been studied. 

At the classical level,  the various NLD equations 
allow for localized solutions with finite energy and charge \cite{ranada:1984}.
This aspect of the NLD equation has led to its use as a model of 
 extended objects in quantum field theory \cite{weyl:1950}. For the  (1+1) dimensional NLD equation (i.e., one 
space dimension plus one time dimension),  analytical solitary wave solutions have been obtained 
for the quadratic nonlinearity \cite{lee:1975,chang:1975}, 
for fractional nonlinearity \cite{mathieuprd:1985} as well as 
for general nonlinearity \cite{stubbe:1986,cooper:2010,xu:2013}.  These results are 
 summarized by Mathieu \cite{mathieu:1985}.
The interaction dynamics of these solitary waves
 has been investigated  in a series of works 
\cite{alvarez:1981,shao:2005,shao:2006,shao:2008,xu:2013,tran:2014} where quite insightful nonlinear phenomena have been found.

One interesting question is how these solitary waves behave when they are placed in trapping potentials.  Whether these solitary waves get trapped  by these potentials or escape freely depends on competition between 
 the effective size of the trapping potential and the 
soliton width.  The effect of  this competition in a spatially periodic parametric force proportional to $\cos(K x)$ has been studied in several systems having soliton solutions:
 the sine-Gordon equation \cite{scharf:1992,sanchez:1992,cuenda:2005},   self interacting  $\varphi^4$ theory, and in  nonlinear Schr\"odinger (NLS) equations \cite{sanchez:1994,scharf:1993}.
The length-scale competition
appears when the soliton width is comparable with  the period $\lambda=2 \pi/K$.
In this situation, the dynamics of the soliton is near a transition from bound to unbound behavior and is very sensitive to the specific details of the nonlinear dynamics. 
     In this paper we will  use the method of collective coordinates in a variational formulation which includes dissipation to study approximately
the response of exact solitary wave solutions of the  NLD equations with arbitrary nonlinearity parameter $\kappa$  to forcing that is proportional to $\cos(K x)$.  The effect of the dissipation on the dynamics of the soliton will also be studied. 

The use of collective coordinates to study solitary waves has a long history. 
In conservative systems, variational  methods were used to study the effect of small perturbations on the solitary waves found in the Korteweg-de Vries (KdV) equation, the modified KdV equation and the NLS equation  \cite{bondeson:1979}. A similar approach was  used to obtain the approximate  time evolution of two coupled NLS equations  \cite{kishvar:1990},  starting with trial wave functions based on the exact solution of the uncoupled problem.  From a  different perspective, Cooper and collaborators were interested in using robust post-gaussian trial wave functions which were {\em not} in the class of exact solutions to understand how well one could approximate exact solutions using this class of functions and to also understand what properties of soliton dynamics did not rely on knowing the exact solution
\cite{cooper:1992, cooper:1993a,cooper:1993b,cooper:1993c,cooper:1994}.   More recently it has been realized that if one chooses a robust enough trial wave function one can obtain a good estimate of the stability of the solitary wave in external environments by studying the linear stability of the reduced set of collective coordinates.  Specifically, this method presumes that the influence of the parametric force on the soliton can be captured by assuming that some shape parameters of the wave function become functions of time. 
An example of this is found in  \cite{cooper:2017} and references therein. 

For the solitary waves of the NLD equation, the question of length scale competition for the  NLD equation with parametric driving of the type  proportional to $\cos(K x)$  was studied in  \cite{quintero2:2019}  using a recently developed five collective coordinates (5CC) approximation \cite{quintero:2019}.  In that paper, which only considered nonlinearity parameter $\kappa=1$ (the interaction term in the Lagrangian density being $(g^2/2) (\bPsi\Psi)^{2}$)  it was shown that the behavior of the collective coordinates agreed quite well with numerical simulations of the various moments.
 It was also found that at  $K=0$, the center of the soliton moved (apart from rapid small oscillations) like a free particle. Once $K$ was turned on, as long as the solitary wave was moving, when   $2 \pi/K > l_s$  the solitary wave gets trapped and the  collective coordinates of the solitary wave oscillate with  two frequencies, the faster one being around  $2 \omega_0$, where $\omega_0$ is the initial frequency of the soliton solution. 
When $2 \pi/K \gg l_s$, the effect of the driving term which acts as a trapping  potential goes to zero and the solitary wave again moves freely with small  rapid oscillations given approximately by $2 \omega_0$.  In the intermediate regime the solitary wave was subject to instabilities at late times that are found in numerical simulations.  This effect was not captured by the 5CC approximation  \cite{quintero2:2019}.

In this paper we consider a generalization of the problem studied in  \cite{quintero2:2019}, where the nonlinearity term in the Lagrangian is taken to an arbitrary power  $\kappa$.  This power controls the width and shape of the solitary wave: as one increases $\kappa$  one decreases the width of the solitary wave. Exact solutions of the NLD equation at arbitrary $\kappa$ and their stability properties were studied
in \cite{cooper:2010}.  We will use the form of these solutions in the 5CC approximation  to study the transition region between trapped and unconfined motion in the presence of parametric driving. 
Because the shape dependence is a function of $\kappa$,  we expect and indeed find that  at fixed value of $K$, and the same initial conditions, one can go from the ``free particle" regime to the trapped regime by increasing $\kappa$. 
When the dissipation is included the soliton loses energy until it disappears.   As we increase $\kappa$ for fixed $\omega_0$, the solitary wave gets more ``spike like" so that it looks closer to a point particle.  As a result of this the domain where the solitary wave is trapped gets enhanced.  We map out this domain using the collective coordinate approach.  We also compare the motion of the solitary wave at $\kappa=1/2$ and $ \kappa=2$ with numerical simulation of the NLD equations and get qualitative agreement.  The transition regime curves would have been impossible to determine from numerical simulations for both numerical reasons and time feasibility constraints.

 The paper is organized as follows.  In Sec. \ref{sec2} both the parametric force and the damping  are introduced in the NLD equation and dynamical equations for the charge, the momentum and the energy are derived. We also show that the equations of motion can be obtained from a Lagrangian when we include a dissipation function \cite{Rayleigh}. In Sec.  \ref{sec3}  an ansatz with five collective coordinates is used as an approximate  solution of the parametrically driven NLD equation and the equations of motion for the collective coordinates are obtained. Then, in Sec. \ref{sec4} we present numerical solutions of the collective coordinates equations and discuss the transition of behavior as a function of $\kappa$ for fixed $K$ as well as the transition of behavior from oscillatory to free for $\kappa=1/2$, $\kappa=1$ and  $\kappa=2$ as we increase $K$.  Finally, the main results of the work and the conclusions are summarized in Sec.  \ref{sec5}.

%%%%%%%%%%%%%%%%%%%%%%%%%%%%%%%%%%%%%%%%%%%%%%%%%%%%%%%%%%%%%%%%%%%%%%%%%%
\section{Damped and parametrically driven NLD equation} \label{sec2}
%%%%%%%%%%%%%%%%%%%%%%%%%%%%%%%%%%%%%%%%%%%%%%%%%%%%%%%%%%%%%%%%%%%%%%%%%%%%
The parametrically driven nonlinear Dirac equation  was recently investigated in Ref.  \cite{quintero2:2019}. Here we generalize this study taking into account that an arbitrary parameter $\kappa$ affects the nonlinear term, so that the parametrically driven and damped NLD equation is given by  
%\red{removed the subscript s below} 
\bq
i \gamma^{\mu} \partial_{\mu}\Psi - m\Psi +g^2 (\bPsi \Psi)^\kappa\Psi 
= f(x) \Psi^\star - i\, \rho\,\gamma^0\,\Psi \>, \label{nlde1}
\eq
where  
\bq \label{q1}
\gamma^{0} = \sigma_{3}= 
\left( \begin{array}{cc}
	1 & 0  \\
	0 & -1  \end{array} \right), \qquad \gamma^{1} = i \sigma_2= %i \sigma^{1}=
\left( \begin{array}{cc}
	0 & 1  \\
	-1 & 0  \end{array} \right),
\eq
$\sigma_2$ and $\sigma_3$ are the Pauli matrices,
%\cite{alvarez:1981}
$\rho$ is the dissipation coefficient, 
and the force reads
\bq
 f(x) = r \cos(K x),  \label{force}
\eq
where $r$ and $K$ are real parameters, which represent the amplitude and the wave number, respectively. Therefore, the inhomogeneous parametric force has a period 
$\lambda=2\pi/K$. The corresponding adjoint NLD equation is
\bq
i  \partial_{\mu} \bPsi \gamma^{\mu} + m \bPsi -g^2 (\bPsi \Psi)^\kappa  \bPsi 
= -f^\star(x) \bPsi^\star - i\, \rho\,\bPsi\,\gamma^0 \>. \label{anlde1}
\eq
The equations (\ref{nlde1}) and (\ref{anlde1}) are derived in a standard fashion from the Lagrangian density
\bq
\mathcal{L} =  \left(\frac{i}{2}\right) [\bPsi \gamma^{\mu} \partial_{\mu}\Psi 
-\partial_{\mu} \bPsi \gamma^{\mu}\Psi] - m \bPsi\Psi 
+ \frac{g^2}{\kappa+1 } (\bPsi\Psi)^{\kappa+1}  \> -\frac{1}{2} f \bPsi\Psi^\star -\frac{1}{2} f^\star \bPsi^\star\Psi, 
\label{laf}
\eq
and from the dissipation function  
\bq
\mathcal{F} =  -i\, \rho\, (\bPsi \gamma^{0} \partial_t\Psi 
- \partial_t \bPsi \gamma^{0}\Psi).
\label{dff}
\eq
Straightforward calculations show that by inserting (\ref{laf}) and (\ref{dff}) into 
\begin{eqnarray} \label{l1}
\partial_{\mu}\,\frac{\partial {\cal L} \quad}{\partial(\partial_{\mu}\bPsi)}-
\frac{\partial {\cal L}}{\partial \bPsi}&=& \frac{\partial \cal{F} \quad}{\partial(\partial_{t}\bPsi)}, \\
\label{l2}
\partial_{\mu}\,\frac{\partial {\cal L} \quad}{\partial(\partial_{\mu}\Psi)}-
\frac{\partial {\cal L}}{\partial\Psi}&=& \frac{\partial \cal{F} \quad}{\partial(\partial_{t}\Psi)},
\end{eqnarray}
Eqs.\ (\ref{nlde1}) and (\ref{anlde1}) are obtained, respectively. 

For a soliton solution of Eq. (\ref{nlde1}), the charge, the momentum and the energy, are, respectively, defined as 
 \begin{eqnarray}  \label{charge}
 Q&=&\int_{-\infty}^{+\infty}dx\,\bPsi\,\gamma^0\,\Psi, 
 \\
 P&=&\frac{i}{2} \int_{-\infty}^{+\infty}dx\, \left( \bPsi_x\,\gamma^0\Psi-\bPsi\,\gamma^0\,\Psi_x
 \right),    
 \label{mome} \\
\label{energy}
 E&=&\int_{-\infty}^{+\infty}dx\,\frac{i}{2} (\bPsi\,\gamma^0\Psi_{t}-\bPsi_{t}\,\gamma^0\,\Psi)-\cal{L} \,. 
 \end{eqnarray}
 Although the dependence of $\kappa$ is not explicitly shown in Eqs.\ \eqref{charge}-\eqref{energy}, the expressions for the charge, the 
 momentum and the energy do depend on $\kappa$ as is immediately shown in the next sections. 
 
Moreover, analogous to what has been studied in \cite{quintero2:2019}, the evolution of the charge, the momentum and the energy are, respectively, given by
\begin{eqnarray} \label{charge1}
\frac{dQ}{dt} &=& -2 \rho Q- i\,  \int_{-\infty}^{+\infty}dx\, [f \bPsi\,\Psi^\star-f^\star \bPsi^\star\,\Psi],  
\\ \label{eqP} 
\frac{dP}{dt} &=& -2\,\rho P+\int_{-\infty}^{+\infty}dx\,[f \bar{\Psi}_x \Psi^\star + f^\star \bar{\Psi}^\star \Psi_x], \\ \label{eqEv}
\frac{dE}{dt}&=&\int_{-\infty}^{+\infty}dx\, \cal{F}. 
\end{eqnarray}

%%%%%%%%%%%%%%%%%%%%%%%%%%%%%%%%%%%%%%%%%%%%%%%%%%%%%%%%%%%%%%%%%%%%%%%%
\section{Five Collective Coordinates ansatz} \label{sec3}
%%%%%%%%%%%%%%%%%%%%%%%%%%%%%%%%%%%%%%%%%%%%%%%%%%%%%%%%%%%%%%%%%%%%%%%%%%%%%
In this section an ansatz is suggested as an approximate solution of Eq.\ (\ref{nlde1}). This solution will depend on time only through the so-called collective coordinates. In particular, for the NLD equation without perturbation, i.e. Eq. (\ref{nlde1}) with $\rho=0$ and 
$f(x)=0$, the two spinor components of a moving soliton solution are 
given by 
\ba \label{eq2.37}
&&\Psi_1(x,t) = \left[ \cosh(\eta/2) A(x') + i \sinh(\eta/2) B(x') \right] e^{-i\omega t'}, \\ \label{eq2.33}
&&\Psi_2 (x,t) = \left[ \sinh(\eta/2) A(x') + i \cosh(\eta/2) B(x') \right] e^{-i\omega t'},
\ea
where $x' = \gamma(x-vt)$, $t' = \gamma(t-vx)$, and 
\ba \label{eq2.37a}
A(x') & = & \sqrt{m+\omega} ~ \bigg [\frac{(\kappa+1) \beta ^2}{g^2} \bigg ]^{\frac{1}{2\kappa}}\, 
  \left[ \frac {1}  {m+\omega \cosh(2\kappa \beta x')} \right] ^{\frac{\kappa+1}{2\kappa }}\,\cosh (\kappa \beta x'),    \\ \label{eq2.33a}
B(x') & = & \sqrt{m-\omega} ~  \bigg [\frac{(\kappa+1) \beta ^2}{g^2}  \bigg ]^{\frac{1}{2\kappa}}\,
 \left[ \frac {1}  {m+\omega \cosh(2\kappa \beta x')} \right] ^{\frac{\kappa+1}{2\kappa }}\,\sinh (\kappa \beta x').   
\ea
In the above equations $\omega$ is a constant frequency, $\gamma=\cosh(\eta)=1/\sqrt{1-v^2}$ is the  Lorentz factor, $\eta$ is the rapidity, and $\beta=\sqrt{m^2-\omega^2}$ \cite{cooper:2010}.  

Due to the  smallness of the perturbation, it is assumed that the only modification to the exact solutions (\ref{eq2.37}), (\ref{eq2.33}) of the  
NLD equation is that the constant parameters and linear time dependent variables become unknown time dependent functions \cite{quintero:2019}. 
This implies that in Eqs.\ (\ref{eq2.37}) and  (\ref{eq2.33})  we replace
\ba
&&vt \rightarrow q(t);~~ \omega \rightarrow \omega(t);  
~~ \eta \rightarrow \eta(t);   \nonumber \\
&&  \omega t'=  \gamma\omega( t - vx) \rightarrow \phi(t) -p(t)(x-q(t)), ~~ x'=\gamma( x-vt) \rightarrow z=(x-q(t) ) \cosh \eta(t). \nonumber
\ea
Thus, our trial wave function, with the five collective coordinates $q(t)$, $\omega(t)$, 
$\eta(t)$, $\phi(t)$, and $p(t)$, reads \cite{quintero:2019}
\begin{eqnarray}
\label{psi1aXX}
\Psi_{1a}(x,t)&=& e^{-i \phi(t)+i p(t) [x-q(t)]} \left\{
\cosh[\eta(t)/2] A(z,t)+i \sinh[\eta(t)/2] B(z,t)
\right\}, \\
\label{psi2aXX}
\Psi_{2a}(x,t)&=& e^{-i \phi(t)+i p(t) [x-q(t)]} \left\{
\sinh[\eta(t)/2] A(z,t)+i \cosh[\eta(t)/2] B(z,t)
\right\},
\end{eqnarray} 
where the variable $z$ will be useful when we perform integrations over all $x$. Moreover, $A(z,t)$ and $B(z,t)$ are given by 
\ba \label{eqAA}
A(z,t)  & = &  \sqrt{ \frac{m+\omega(t)}{m+\omega(t)  \cosh(2\kappa \beta(t) z)}} 
\bigg \{\frac{(\kappa+1) \beta(t)^2  }{g^2 [m+\omega(t)   \cosh (2\kappa \beta(t) z)]} 
\bigg \}^{\frac{1}{2\kappa}} \, \cosh [\kappa \beta(t)  z],    \\ \label{eqBB}
B(z,t) & =  &  \sqrt{ \frac{m-\omega(t) }{m+\omega(t) \cosh(2\kappa \beta(t) z)}} 
\bigg \{\frac{(\kappa +1) \beta(t) ^2}{g^2 [m+\omega(t)  \cosh (2\kappa \beta(t) z)]} 
\bigg \}^{\frac{1}{2\kappa}}\,\sinh(\kappa \beta(t) z), 
\ea
where now $\beta(t)=\sqrt{m^2-\omega^2(t)}$ is a time dependent function. 

Inserting (\ref{psi1aXX})-(\ref{psi2aXX}) into (\ref{laf}) and integrating over $x$  we determine the effective Lagrangian for the variational parameters.
For $L$ we have
\ba \label{la2XX}
L(\omega,p,\phi,\eta, q, \dot{q},\dot{\phi}) &=& \int {\cal L} dx  = Q(\omega)\,\left[p \dot{q}+\dot{\phi}-p\,\tanh \eta \right]-  \\ 
&-& I_0(\omega)[\cosh \eta-\dot{q} \sinh \eta]-\frac{\omega\,Q(\omega) }{\cosh \eta}-U(\omega,p,\phi,\eta,q). \nonumber 
\end{eqnarray} 
Here $I_0(\omega)$ is represented by Eq.\ \eqref{I0} (see appendix \ref{appendix}), and the charge $Q(\omega)=\int dx \Psi^\dag \Psi=\int dx (|\Psi_{1a}|^2+|\Psi_{2a}|^2)$ is  calculated by using the ansatz \eqref{psi1aXX}-\eqref{psi2aXX}, and is given by 
\bq \label{Q}
Q(\omega) =  \frac{1}{  \kappa \beta }   \left[ \frac{(\kappa+1) \beta^2}{g^2  (m+\omega )}\right]^{1/(\kappa)} I_\kappa[\alpha^2,\kappa] \,, 
\eq
where
\ba \label{Ik} 
&& I_\kappa[\alpha^2, \kappa]=B(1/2,1/\kappa) _2F_1(1+1/\kappa,1/2,1/2+1/\kappa;\alpha^2) +\alpha^2 B(3/2,1/\kappa) _2F_1(1+1/\kappa,3/2,3/2+1/\kappa;\alpha^2).
\ea
Here $B(x,y)$ and $_2F_1$ denote Beta functions and hypergeometric functions, respectively.  The calculations to obtain the above expression for the charge are similar to the ones presented in \cite{cooper:2010} and for this reason are omitted.
The effective potential reads (see appendix \ref{appendix})
 \ba \label{Ukappa}
&&U 
= \frac{ r}{\cosh \eta}  \left(\frac{  (\kappa+1) \beta ^2}{g^2 \omega}\right)^{1/\kappa}    \\
&& \times \left\{ \cos( K q(t) + 2 \phi(t)) \left[\frac{m}{ \omega} I_p[a,b_{-} ,c,\nu] - \frac{\beta^2}{\omega^{2}}  I_p[a,b_{-} ,c,\nu+1]  \right]  \right.   \nonumber \\
&& \left. +\cos( K q(t) - 2 \phi(t)) \left[\frac{m}{ \omega} I_p[a,b_{+} ,c,\nu] - \frac{\beta^2}{\omega^{2}}  I_p[a,b_{+} ,c,\nu+1]  \right] \right\} \,, 
\ea
where $I_p[a,b,c,\nu]$, $a$, $b$, $c$ and $\nu$ are parameters defined in appendix \ref{appendix}.
%%%%%%%%%%%%%%%%%%%%%%%%%%%%%%%%%%%%%%%%%%%%%%%

%%%%%%%%%%%%%% AAAQQQUUUUIIIIIII

   Note that $U$ explicitly depends on $q(t)$ when $K \neq 0$ so there is now a force coming from the driving term.   Momentum is no  longer conserved in the presence of this type of forcing which breaks the  parity symmetry once $q(t)$ is not zero.

%In general, we have from Eqs. \eqref{t11}, \eqref{m0}, that
%\bq
%\omega  Q - H_2 + H_1/ \kappa =0,
%\eq
%Therefore we can replace $H_1$ by
%\bq
%H_1 = \kappa(H_2- \omega Q) 
%\eq
% 
%\bq \label{m0}
%M_0 = \int dx  h_1\left(1-\frac{1}{\kappa}\right)+h_2\,= H_1\left(1-\frac{1}{\kappa}\right)+H_2\,
%\eq
%Thus at $\kappa=1$ we get the relations:
%\bq
%M_0 = H_2;~~ I_0= H_1=M_0-\omega Q
%\eq
% 
% 
%For $\kappa=1$, $U, Q,M$, and $I_0$ simplify and are given by 
%
%\begin{eqnarray} 
%%L(\omega,p,\phi,\dot{q},\dot{\phi},\eta)&=& Q(\omega)\,\left[p \dot{q}+\dot{\phi}-\frac{\nu}{2}-\left(p-\frac{k}{2}\right)\,\tanh \eta \right]- \nonumber \\ \label{la2XX}
%%&-& I_0(\omega)[\cosh \eta-\dot{q} \sinh \eta]-\frac{\omega\,Q}{\cosh \eta}-U(\omega,p,\phi,\eta), \\ \label{la3XX} 
%U(\omega,p,\phi,\eta)&=&\frac{r\,\cos(2\,\phi)}{\cosh \eta}\,
%Q\,\frac{a\,\pi }{\sinh a \pi}\,\cos(a\,\cosh^{-1}m/\omega),
% \quad Q(\omega)= \frac{2\,\beta}{g^2\,\omega}, \\ \label{la5XX}
% M_0(\omega)&=&\frac{4\,m}{g^2}\,\tanh^{-1}(\alpha),  \alpha=\sqrt{\frac{m-\omega}{m+\omega}}, a=\frac{p}{\beta\,\cosh \eta},  \beta=\sqrt{m^2-\omega^2}. \nonumber \\.
%\end{eqnarray}
%Moreover, $I_0(\omega)=M_0(\omega)-\omega Q(\omega)$. From now on is assumed that $\mu=0$. 
%
% 
By inserting (\ref{psi1aXX})-(\ref{psi2aXX}) into (\ref{dff}) and integrating over $x$  we can calculate the dissipation function $F$ for the collective coordinate equations. 
We  obtain
\bq \label{DF}
F = -2 \rho  \left[ I_0 \sinh \eta {\dot q} + Q (p \dot q + \dot \phi) \right].
\eq
We see that the effect of dissipation is to modify the equation for the momentum conjugate to $q$  and the charge $Q$ which is 
conjugate to $\phi$. 
%%%%%%%%%%%%%%%%%%%%%%%%%%%%%%%%%%%%%%%%%%%%%%%%%%%%%%%%%%%%%%%%
\section{Lagrange Equations for the Collective Coordinates } \label{sec4}
%%%%%%%%%%%%%%%%%%%%%%%%%%%%%%%%%%%%%%%%%%%%%%%%%%%%%%%%%%%%%%%%%%%

From the Lagrangian \eqref{la2XX} 
and the dissipation function Eq. {\eqref{DF}, 
we obtain the equations of motion for the collective coordinates using the Lagrange equations and Rayleigh's dissipation functional formalism \cite{Rayleigh}.  
%\bq \label{lqXX}
%\frac{d}{dt} \frac {\partial L}{\partial \dot q_i} -  \frac{\partial L} {\partial q_i} = 0,
%\eq
We obtain the canonical momentum conjugate to $q(t)$ as 
\begin{equation} \label{Pq}
P_q=\frac{\partial L}{\partial \dot{q}}=Q\,p+I_0 \sinh \eta, 
\end{equation}
which obeys the equation
\bq \label{Pdot}
\dot P_q =  -2 \rho P_q - \frac{\partial U}{\partial q}. 
\eq

This leads to the equation of motion:
\bq  \label{eq2}  
Q \dot p + p \frac{dQ}{d\omega} \dot \omega + I_0 \cosh \eta \dot \eta + \frac{dI_0}{d\omega}  \sinh \eta \dot \omega= - \frac{\partial U}{\partial q}  -2 \rho P_q. 
\eq
The Lagrange equation obtained by choosing  $ q_i = \phi$ reads 
\begin{eqnarray} 
& &\frac{dQ}{dt}= \frac{dQ}{d \omega}  \dot \omega = -\frac{\partial U}{\partial \phi}- 2 \rho  Q.\label{relphiXX}
\ea
This equation can be written as 
\bq \label{eq4}
 \dot \omega = -\dfrac{ 2 \rho  Q+ \dfrac{\partial U}{\partial \phi}} {dQ/d \omega}.
 \eq
Choosing $q_i= p$ we obtain
 % \red {This is Niurka eq1}
 \ba
\label{eq1a}
& &Q\,(\dot{q}-\tanh \eta)=\frac{\partial U}{\partial p},  \label{omega}
\ea
which leads to the first-order equation:   
\bq  \label{eq1}
\dot q = \tanh \eta + \frac{1}{Q}  \frac{\partial U}{\partial p} \,. 
 \eq
Choosing $q_i=\omega$ we get
 \ba     \label{phidot}
\left[p \dot{q}+\dot{\phi}- p  \tanh \eta -\frac{\omega}{\cosh \eta}\right] \frac{dQ}{d\omega}- \frac {Q} {\cosh \eta}    
-\frac{dI_0}{d \omega} [\cosh \eta - \dot q \sinh \eta]
=\frac{\partial U}{\partial \omega} \,.    \nonumber 
\ea
This equation can be simplified using Eq. \eqref{rel1}, namely that $ \frac{dI_0}{d \omega} +Q(\omega) =0$, which  leads to
\ba
& &\left[p \dot{q}+\dot{\phi}- p  \tanh \eta -\frac{\omega}{\cosh \eta}\right] \frac{dQ}{d\omega}-Q \sinh \eta (\dot{q}-
 \tanh \eta) =\frac{\partial U}{\partial \omega}.\label{relQXX} 
 \ea
 Using  Eq. \eqref {eq1}, we can eliminate $\dot q$  to obtain:
 \ba
& &\left[\dfrac{p}{Q} \dfrac{\partial U}{\partial p} +\dot{\phi} -\frac{\omega}{\cosh \eta}\right] \frac{dQ}{d\omega}- \sinh \eta  \frac{\partial U}{\partial p} =\frac{\partial U}{\partial \omega}. \label{relQXXX} 
 \ea
The equation for $\phi(t)$ reads
 \ba \label{eq3} 
 && \dot \phi  =  \frac{\omega}{\cosh \eta} - \frac{p}{Q}  \frac{\partial U} {\partial p}    
 + \sinh \eta \frac{ \frac{\partial U} {\partial p}}{ \frac{\partial Q}{\partial \omega}} +\frac{ \frac{\partial U} {\partial \omega }}{ \frac{\partial Q}{\partial \omega}}. 
 \ea
 Choosing $q_i=\eta$ we obtain 
\ba
& & \left(\omega \sinh \eta -p \right) \frac{Q}{\cosh^2 \eta}- {I_0}(\sinh \eta-\dot q \cosh \eta) =\frac{\partial U}{\partial \eta}.   \nonumber \\ 
\end{eqnarray}
Using the equation for $\dot q$ we can reduce this to 
an algebraic equation relating the five collective coordinates: 
\ba \label{eq5} 
& & \left(\omega \sinh \eta -p  \right) \frac{Q[\omega] }{\cosh^2 \eta}+{I_0[\omega]}\frac{1}{Q[\omega] }  \frac{\partial U[p,\omega,\eta,\phi,q]}{\partial p}   \cosh \eta =\frac{\partial U[p,\omega,\eta,\phi,q]}{\partial \eta}.   
\end{eqnarray}
This nonlinear algebraic equation Eq. \eqref{eq5}  allows one to determine $p(0)$ in terms of  $\omega_0,\eta_0,\phi_0$ and $q_0$ at $t=0$. 

So we have five coupled equations to solve.  Four are ODE's for $\omega$, $q$, $\phi$, and $\eta$, namely Eqs. \eqref{eq1}, \eqref{eq2}, \eqref{eq3}, \eqref{eq4} and then an algebraic equation for $p$, namely Eq. \eqref{eq5}.
The condition for collapse of the wave function is that $\beta \rightarrow 0$, which happens when there is dissipation.

Notice that from \eqref{Pq} and \eqref{la2XX}, we obtain 
\ba \label{la2XXX}
L(\omega,p,\phi,\eta, q, \dot{q},\dot{\phi}) = P_q\, \dot{q}\,+Q(\omega)\,\dot{\phi}-E, 
\end{eqnarray}
where the energy $E$ is obtained by inserting the ansatz in Eq.\ \eqref{energy}. This procedure yields
\ba \label{energy1}
E = P_q\, \tanh \eta +\frac{M_0}{\cosh \eta} +U,  
\end{eqnarray}
where $M_0=I_0+\omega\,Q$. 
It is also worth mentioning that by inserting the ansatz in Eq.\ \eqref{mome}, we obtain  that the field momentum is equal to the canonical momentum $P(t)=P_{q}(t)$. 

In our previous paper \cite{quintero2:2019}  
which discussed the case $\kappa=1$ we showed that these 5 CC equations satisfy exactly the  moment equations for  the evolution of charge $Q$  \eqref{charge1}, momentum  $P$ \eqref{eqP}, energy $E$  \eqref{eqEv} (which leads to 2 CC equations)  and first moment of the charge.

\subsection{No force, and no damping}

Equation \eqref{Pdot} implies $P_q$ dissipates exponentially with $t$ in the absence of external forcing but when the dissipation $\rho$ is non-zero. From the equations of motion, we recover the exact soliton solution when $U=0$,  and $\rho=0$. First, from Eq. \eqref{eq4} $\omega=\omega_0$, a constant is obtained. Second,  \eqref{eq1} leads to the usual connection between $\dot q$ and $\eta$, namely
$\dot{q} = \tanh \eta$. Third, Eq. \eqref{eq3} leads to 
\ba \label{dotphizero}
\dot \phi = \dfrac{\omega}{\cosh \eta}.
\ea
Moreover, $U=0$ in \eqref{eq5} implies  
$
p=\omega \sinh \eta
$, 
and that  $\eta$ and $\omega$ are constant so that by integrating Eq. \eqref{dotphizero} we obtain
\bq \phi(t) = \frac{\omega t}{\cosh \eta}. \eq
 Finally, the total  phase is 
 \bq
 \phi(t) -p(t) (x-q(t)) \equiv \omega t'= \omega \cosh \eta (t - v x),
\eq
where $v= \dot q = \tanh \eta$.
Thus in this case the trial wave function becomes the \emph{exact} wave function.

%%%%%%%%%%%%%%%%%%%%%%%%%%%%%%%%%%%%%%%%%%%%%%%%
\section{Numerical study}
%%%%%%%%%%%%%%%%%%%%%%%%%%%%%%%%%%%%%%%%%%%%%%%%
Numerical simulations of the parametrically driven NLD equation (\ref{nlde1}) have been performed in order to check the validity of the approximate solution (\ref{psi2aXX}), with the five collective coordinates $q(t), \omega(t), \eta(t), \phi(t)$ and $p(t)$.  For these simulations, we have taken as initial condition the ansatz (\ref{psi2aXX}) evaluated at $t=0$, by specifying the initial values  $q(0),\omega(0), \eta(0), \phi(0)$,  and $p(0)$, taking into account  that $p(0)$ is determined  by  the algebraic equation~(\ref{eq5}). In all the simulations we have fixed $m=1$, $g=1$, $\omega_0=0.9$ and $q(0)=0$. A Runge-Kutta-Verner algorithm with variable time step and a spectral method for the spatial derivatives have been employed, and periodic boundary conditions have been set. The system has been discretized by taking $N=3200$ points separated by a constant spatial interval $\Delta x=0.02$. Due to the periodic boundary conditions, the system length $L=N \Delta x=64$ has to be an integer multiple of the spatial period $\lambda=2 \pi/K$, that is $L=n\lambda$. Additionally, $\lambda$ has to be commensurable with $\Delta x$, that is $\lambda =j \Delta x$. Both conditions imply $N=n j$, with $n$ and $j$ being integers. Since $N=3200=2^7 \cdot 5^2$, the possible values of $n$ and $j$ in the simulations are restricted to multiples of $2$, $5$, $25$ or products of them.

%%%%%%% Figure 1%%%%%%%%%%%%%%%%%%%
\begin{figure}
	\begin{tabular}{c}
		\includegraphics[width=0.4\linewidth]{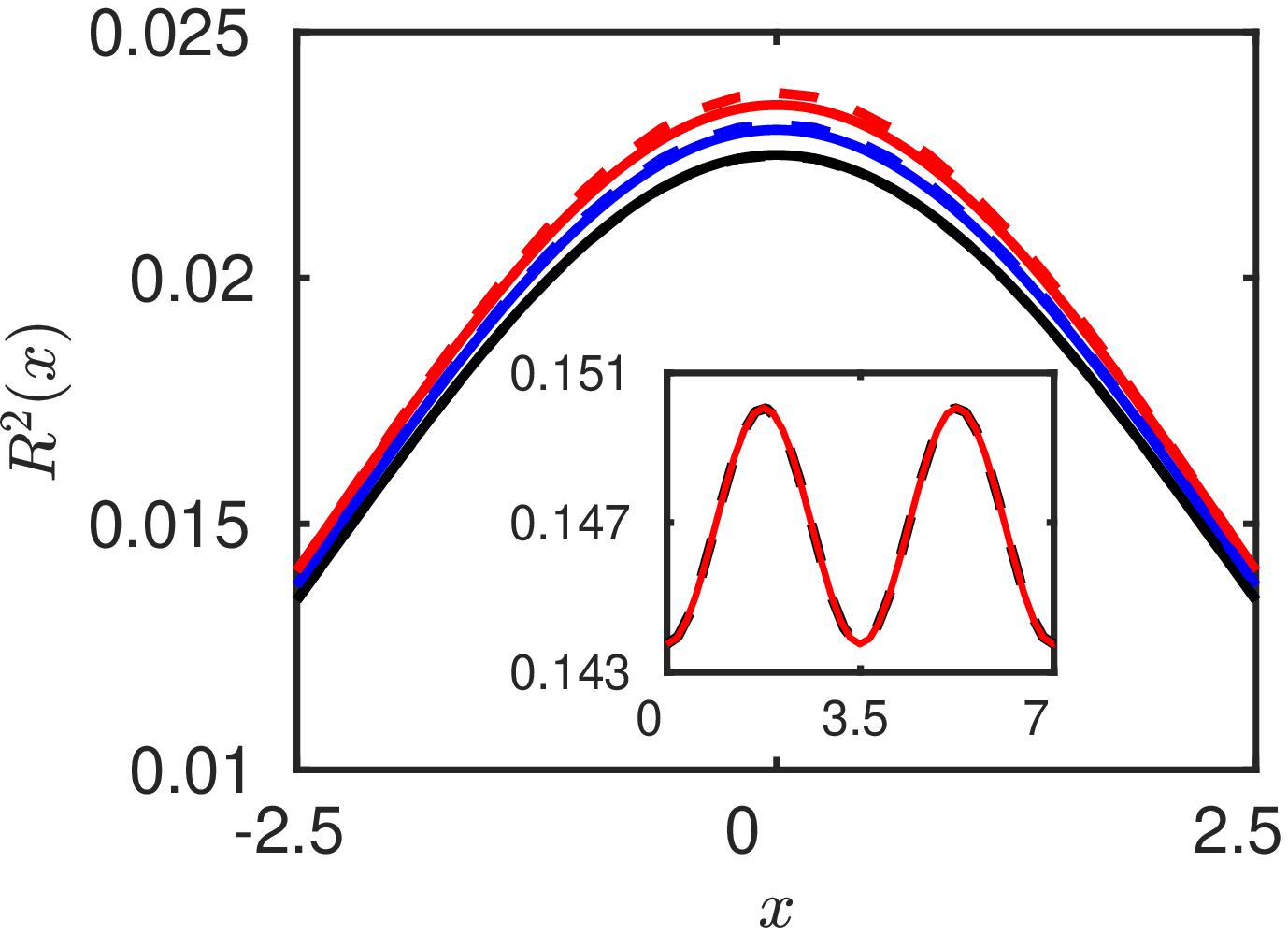} \\
		\includegraphics[width=0.4\linewidth]{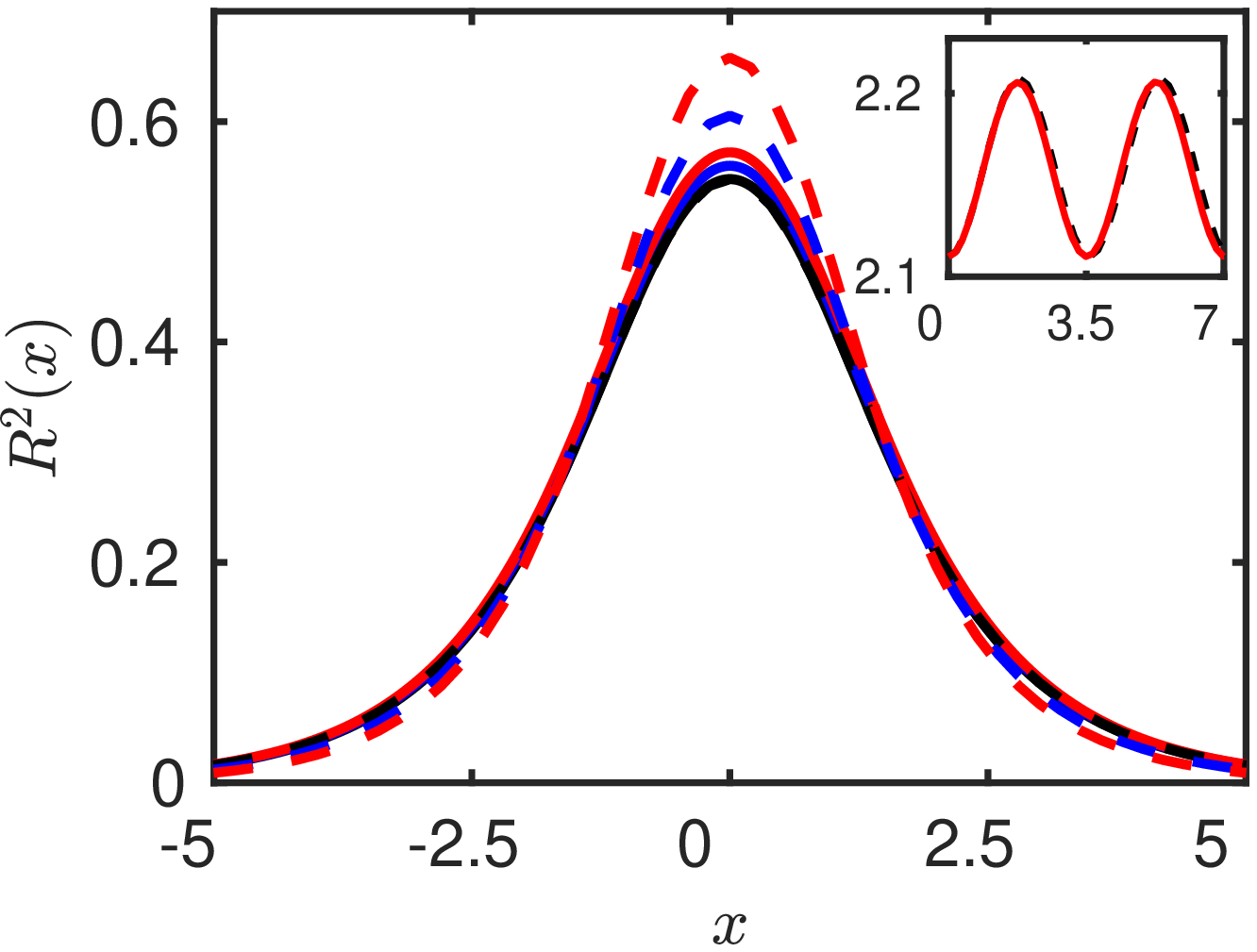}
	\end{tabular}
	\caption{Oscillations of the charge density of a static soliton for $t_0=0, t_1=T/4$ and $t_2=T/2$ (black, blue and red solid line, respectively), with $T=\pi/\omega(0)$ being the period of the oscillations.
		Upper panel: $\kappa=1/2$. Lower panel: $\kappa=2$. The black, blue and red dashed lines correspond to results from the collective coordinate theory. In the inset, the charge oscillations in time are shown (simulations with black dashed line, and collective coordinate results with red solid line). Parameters: $m = 1$, $g = 1$, $r = 0.02$, and $K = \pi/32$. Initial conditions: $q(0) = 0$, $\omega(0) = 0.9$, $\phi(0) = 0$, $\eta(0) = 0$, and $p(0) = 0$.
		\label {fig1}}
\end{figure}
%%%%%%% END Figure 1%%%%%%%%%%%%%%%%%%%

%%%%%%%  Figure 2%%%%%%%%%%%%%%%%%%%
\begin{figure}
	\begin{tabular}{c}
		\includegraphics[width=0.4\linewidth]{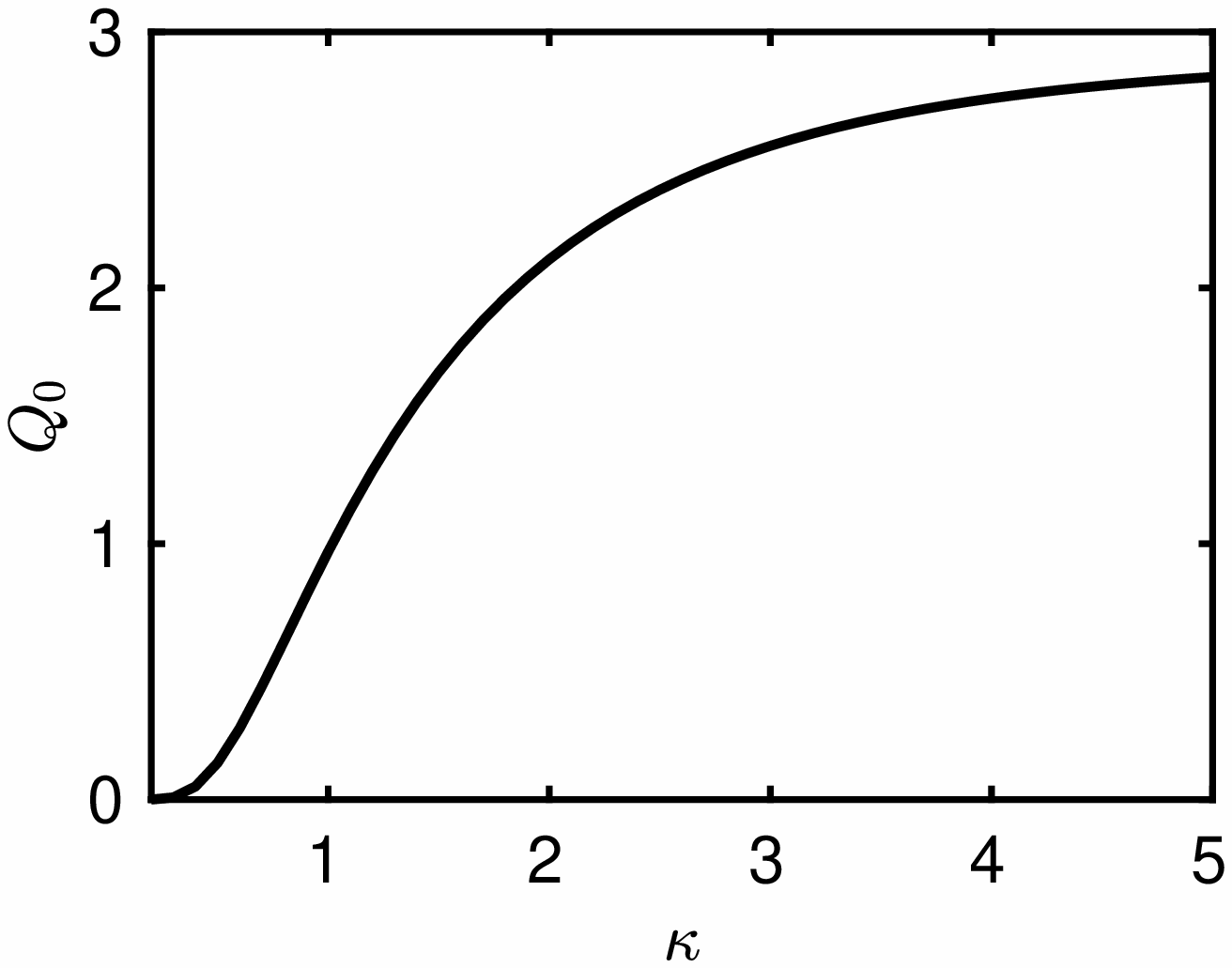}  \\
		\includegraphics[width=0.4\linewidth]{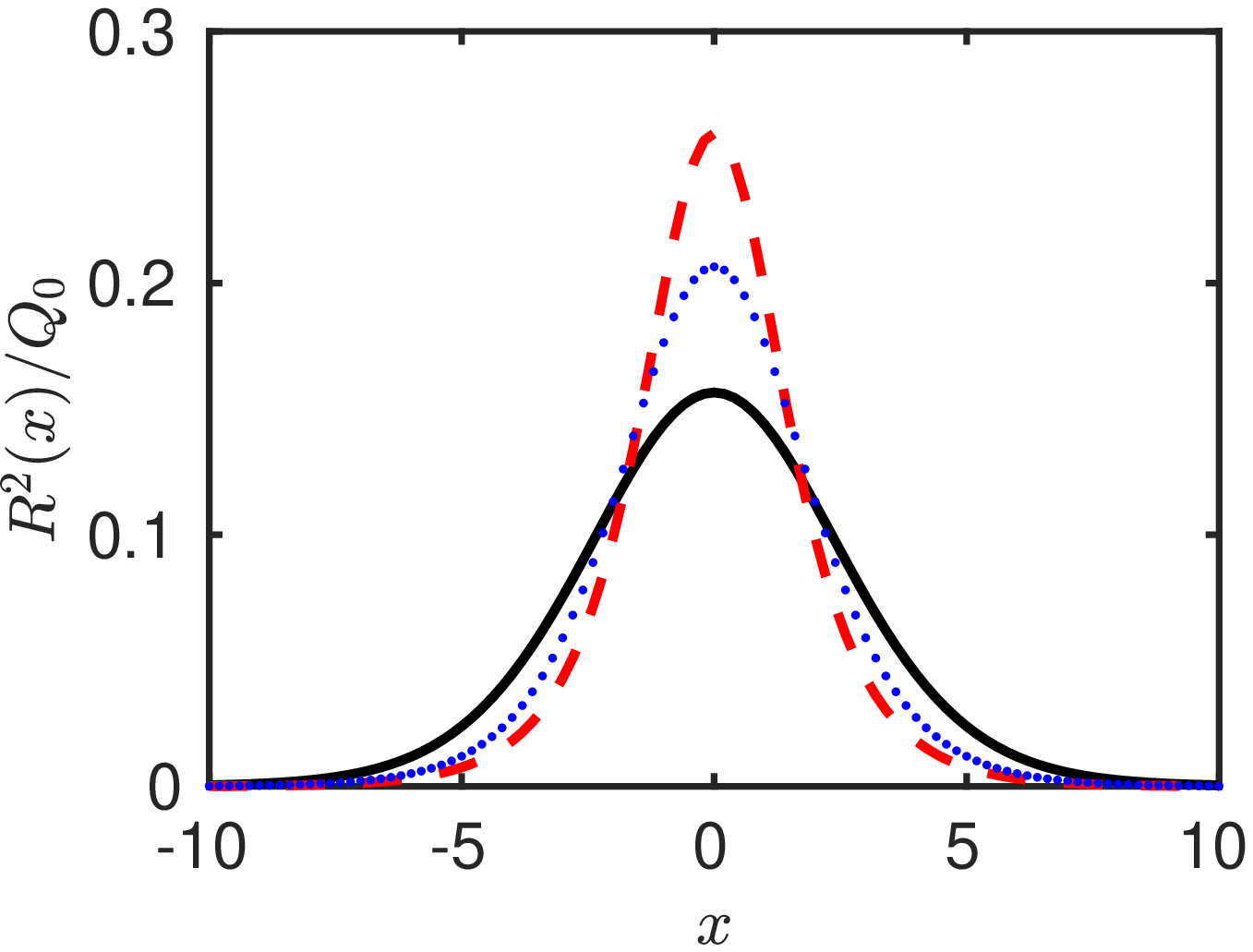}
	\end{tabular}
	\caption{\label{fig2} Top panel: initial charge  versus $\kappa$. Bottom panel: Normalized initial density of charge for $\kappa=1/2$ (black solid line), $\kappa=1$ (blue dotted line) and $\kappa=2$ (red dashed line). For both panels $m = 1$, $g = 1$ and $\omega(0)=0.9$. }
\end{figure}
%%%%%%% END Figure 2%%%%%%%%%%%%%%%%%%%

When there is no dissipation in the system, $\rho=0$, the energy of the soliton is a conserved quantity as seen from Eq. \eqref{eqEv}. For a first test, we have considered the special case of a solitary wave at rest  $\eta(t)=p(t)=0$, and centered at the origin $q(t)=0$. In this case, the solitary wave obeys at all times the equations:
\bq
\dot \phi = \omega  + \frac{ \frac{\partial U} {\partial \omega }}{ \frac{\partial Q}{\partial \omega}},
\eq
and
\bq
\dot \omega = \frac{ -  \frac{\partial U}{\partial \phi}} {dQ[\omega]/d \omega},
\eq
where in the potential function $U$, we have set $p=q=\eta=0$. As a consequence, $U$ becomes independent of the parameter $K$.
%Specifically one has
%\ba
%&&U[p=q=\eta=0] = \nonumber \\
%&& \frac{\sqrt{\frac{\pi }{2}} r \Gamma \left(1+\frac{1}{\kappa }\right) \cos (2 %\phi )
%   \left(\frac{\omega }{m+\omega }\right)^{\frac{1}{\kappa }} %\left(\frac{(\kappa +1)
%       (m-\omega ) (m+\omega )}{g^2 \omega }\right)^{\frac{1}{\kappa }}}{\kappa  %\omega ^2
%   \sqrt{\omega ^3 (m-\omega )}} \times \nonumber \\
%&&\left[\kappa  m \omega ^2 \,
%_2\tilde{F}_1\left(\frac{1}{2},\frac{1}{2};\frac{1}{2}+\frac{1}{\kappa %};\frac{\omega
%   -m}{2 \omega }\right)-\sqrt{\omega  (m-\omega )} \sqrt{\omega ^3 (m-\omega )} %\,
%_2\tilde{F}_1\left(\frac{1}{2},\frac{1}{2};\frac{3}{2}+\frac{1}{\kappa %};\frac{\omega
%   -m}{2 \omega }\right) \right] .\nonumber \\
%\ea

Although the soliton is at rest, it is found that its charge oscillates in time with frequency $ 2 \omega_0$.  This behavior is shown in figure \ref{fig1}, where soliton profiles have been plotted at $t_0=0$ (black solid line), $t_1=T/4$ (blue solid line), and $t_2=T/2$ (red solid line), with $T= \pi/\omega_0$ being the period of oscillations. For $\kappa=1/2$ (top panel), the collective coordinates approximation (dashed lines) captures very well the soliton profiles, as well as the evolution in time of the charge displayed in the inset. However, for $\kappa=2$ significant deviations appear around the maxima of the charge density, although the collective coordinates theory continues to  fit well the charge oscillations (see the inset).  Interestingly, the amplitude of the charge oscillations depends on the nonlinearity parameter $\kappa$, but the oscillation frequency only depends on the choice of $\omega_0$.

The initial soliton charge grows fast and monotonically as the nonlinear parameter $\kappa$ is increased. This feature  can be clearly appreciated in the top panel of figure \ref{fig2}. On increasing $\kappa$, the soliton's amplitude becomes larger but also its width becomes narrower as shown in the bottom panel, where normalized initial density of charge has been plotted for $\kappa=1/2$ (black solid line), $\kappa=1$ (blue dotted line) and $\kappa=2$ (red dashed line).

%%%%%%%%%%%%%%% FIGURE 3 %%%%%%%%%%%%%%%%%%%%%%%%%
\begin{figure}%[h!]
	\begin{center}
		\includegraphics[width=0.4\linewidth]{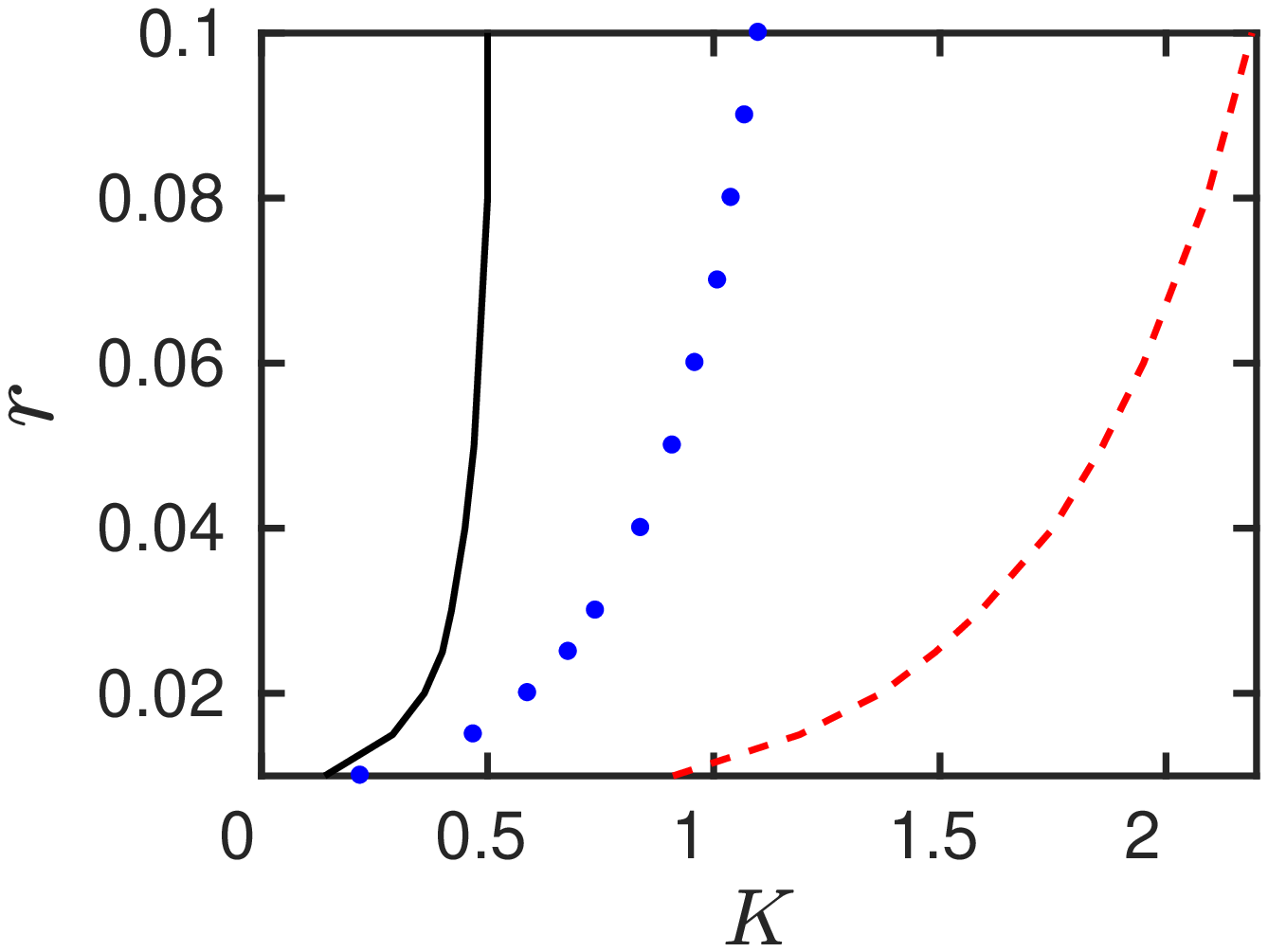}%figrcritK1.eps
	\end{center}
	\caption{Transition from oscillations to soliton net motion for  $\kappa=1/2$ (black solid line), $\kappa=1$ (blue dotted line) and $\kappa=2$ (red dashed line). For a given $\kappa$, the region at the left-hand side of the corresponding line represents oscillating states, while the region at the right-hand side represents free soliton motion. This result has been  obtained using the
		collective coordinates theory for the initial conditions: $q(0)=0$, $\omega(0)=0.9$, $\phi(0)=0$, $\eta(0)=0.01$. The value of $p(0)$ is obtained by solving the algebraic equation (\ref{eq5}).}
	\label{fig3}
\end{figure}
%%%%%%%%%%%%%%% END FIGURE 3 %%%%%%%%%%%%%%%%%%%%%%%%%

In order to obtain a mobile soliton, it is necessary to give it an initial rapidity $\eta(0) \ne 0$.  In the limiting cases $K=0$ (constant parametric force) and $K \gg 1$ ($\lambda$ much smaller than the soliton width), the soliton behaves as a free particle with constant energy and momentum. The width of the soliton does not play any particular role in the dynamics and consequently  its velocity does not depend on the $\kappa$ value.

When $K \ll 1$, the spatial period of the parametric force  $\lambda=2 \pi/K$ is much larger than the width of the soliton. As a consequence, the soliton oscillates inside an effective potential and a second slow frequency, which depends on both $K$ and $\kappa$, modulates all the collective variables. On increasing $K$,  a length scale competition appears between the width of the soliton and the spatial period of the parametric force.  As was shown in a previous paper  \cite{quintero2:2019}, this length scale competition destabilizes the soliton, giving raise to a transition from trapped motion to solitary waves that move like a free particle. Here, we focus on the influence of the nonlinearity parameter $\kappa$ on that transition.

In figure \ref{fig3}, the border that separates soliton oscillations from unbounded motion have been plotted for  $\kappa=1/2$ (black solid line), $\kappa=1$ (blue dotted line) and $\kappa=2$ (red dashed line). These lines have been computed using the collective coordinates approximation, because it allows to vary smoothly the value of $K$ for a fixed amplitude $r$. For a given $\kappa$, the region at the left-hand side of the corresponding line represents oscillating states, while the region at the right-hand side represents unbounded soliton motion. Note that for fixed $r$, the larger $\kappa$ is, the larger $K$ values are needed in order to achieve untrapped soliton motion. This is a direct consequence of the fact that the soliton width decreases with $\kappa$. Therefore, for instance, one should expect a transition from  free motion to oscillations at $r=0.02$ and $K=5 \pi/32 \simeq 0.49$ as we go from $\kappa=1/2$ to $\kappa=2$.  This is indeed observed in the time evolution of $q(t)$ as shown in figure \ref{fig4}.

%%%%%%%%%%%%%%% FIGURE 4 %%%%%%%%%%%%%%%%%%%%%%%%%
\begin{figure}
	\begin{tabular}{c}
		\includegraphics[width=0.4\linewidth] {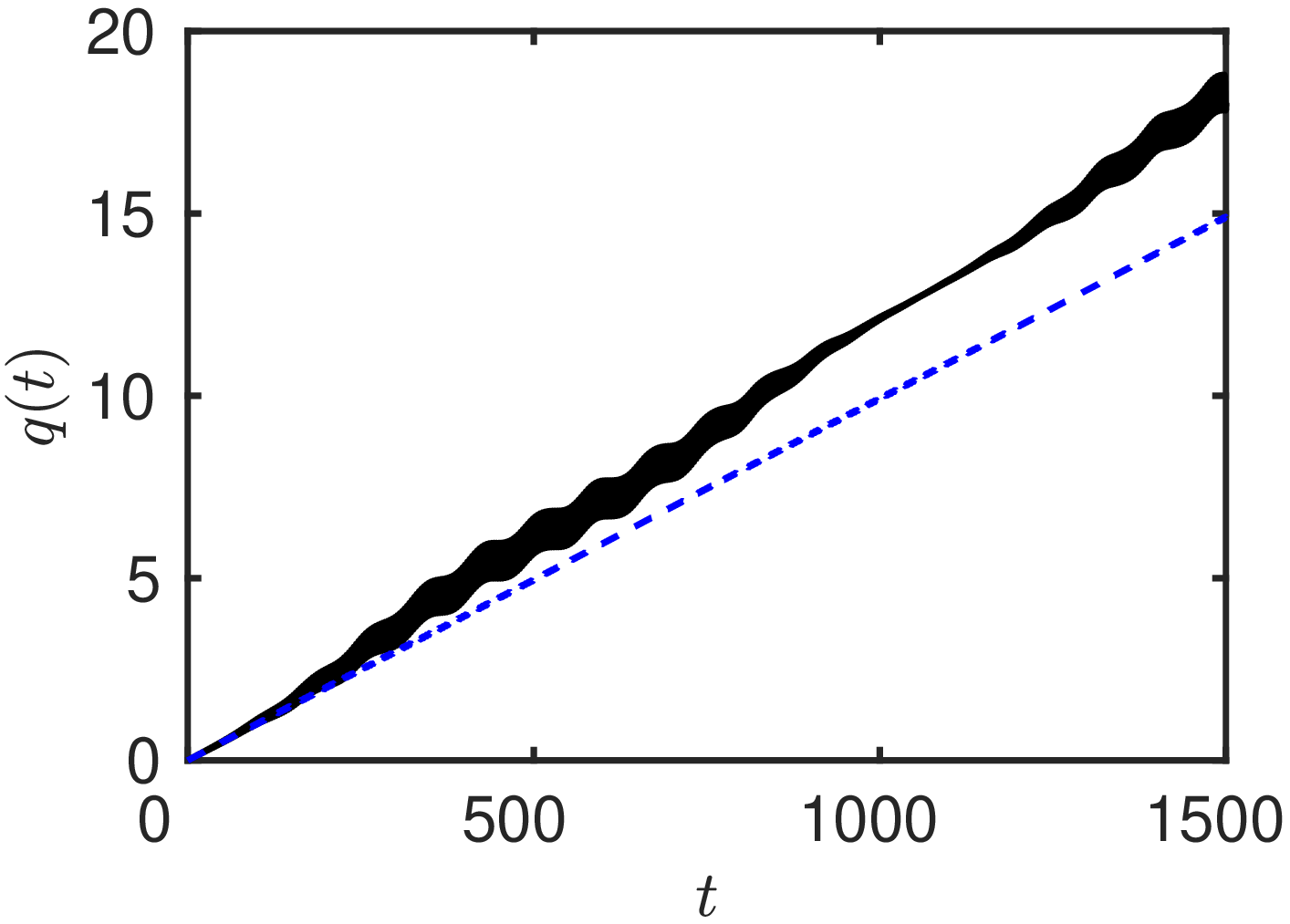}
		\\
		\includegraphics[width=0.4\linewidth] {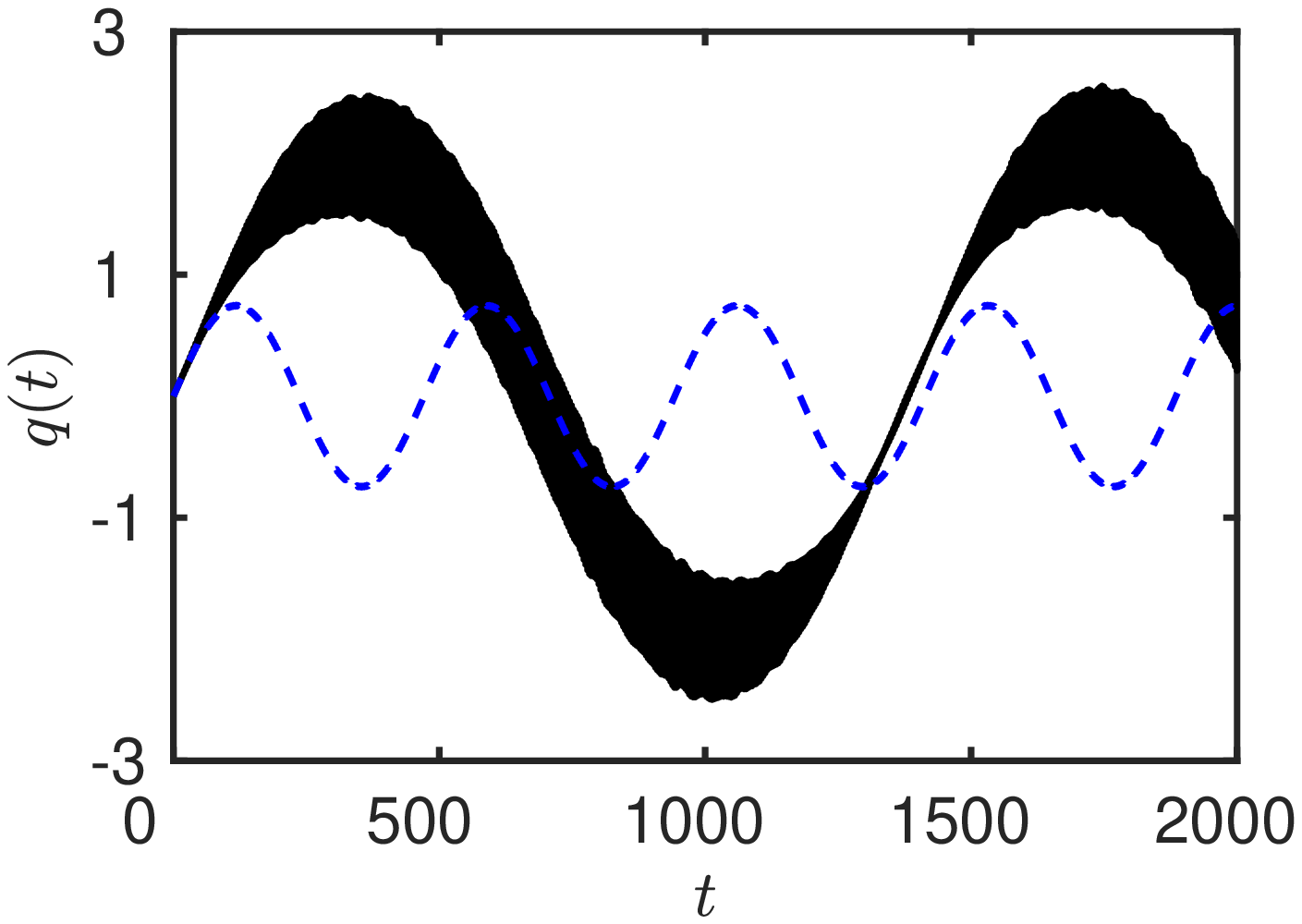}
	\end{tabular}
	\caption{\label{fig4} Motion of the soliton center $q(t)$, for  $K=5\pi/32$, and $r=0.02$. The black continuous line represent simulations of the NLD equation  (\ref{nlde1}) while the blue dashed line corresponds to the collective coordinate approximation. In the top panel,  $\kappa=1/2$ and the motion is unbound. In the  bottom panel, $\kappa=2$ and there is oscillatory trapped behavior. Initial conditions: $q(0)=0$, $\omega(0)=0.9$, $\phi(0)=0$, $\eta(0)=0.01$, and $p(0)=0.00901283$ if $\kappa=1/2$ or $p(0)=0.00904866$  if $\kappa=2$.}
\end{figure}
%%%%%%%%%%%%%%% END FIGURE 4 %%%%%%%%%%%%%%%%%%%%%%%%%

In the top panel, $\kappa=1/2$ and the motion is unbound. In the simulations of the NLD equation  (\ref{nlde1}) represented by a black solid line, very fast oscillations with variable amplitude give raise to changes in the thickness of the line. The collective coordinate approach (blue dashed line) captures well the short time behavior but not the amplitude variations of the fast oscillations.

In the bottom panel of figure \ref{fig4}, $\kappa=2$ and the soliton becomes trapped. There are
significant discrepancies between the collective coordinate theory and the simulations in the amplitude and in the frequency of the slow and large oscillations.

When dissipation  is included in the system, i.e. $\rho \ne 0$, the charge and the energy of the soliton decay exponentially, regardless of the value of $\kappa$ and
$K$.

 \section{conclusions} \label{sec5}
 
We investigated the dynamics of solitary waves in the nonlinear Dirac equation with scalar-scalar self-interaction with arbitrary nonlinearity parameter $\kappa$ and a parametric driving term of the form $r \cos(K x)$. We used a variational approach with five collective coordinates (5CC). The resulting four ODEs plus one algebraic equation were solved numerically by a Mathematica program. The solutions are periodic in time which means that the solitary waves exhibit intrinsic oscillations, plus oscillations in the translational motion. These results were compared with simulations, i.e. numerical solutions of the driven nonlinear Dirac equation.  
 
For a soliton at rest its profile depends strongly on $\kappa$. For $\kappa = 1/2$ the CC approximation captures very well the soliton profile (charge density), as well as the time evolution of the charge. However, for $\kappa = 2$ there are significant deviations in the time evolution of the  charge density when comparing the CC approximation with the numerical solution. Nevertheless, the total charge $Q(t)$  oscillations are described well by the CC theory (figure \ref{fig1}) when compared with numerical simulations.
 
For a moving soliton in the case $K \gg 1$ (spatial period $\lambda = 2 \pi/K$ of the parametric force much smaller than the soliton width $l_s$) the soliton behaves as a free particle with constant charge, momentum and energy. The velocity does not depend on $\kappa$. For $K \ll 1$, $\lambda$ is much larger than $l_s$.  Here the soliton oscillates inside an effective potential with frequency $2 \omega(0)$, where $\omega(t)$  is a CC, and a second frequency, which is very low and depends on both $K$ and $\kappa$, modulates all collective coordinates.
 
Increasing K, a length scale competition appears between $l_s$ and $\lambda$. In a plot of the amplitude $r$ of the driving force vs. $K$ there is a border line that separates the trapped motion from the free motion of the soliton. This border line depends very strongly on the value of $\kappa$ (figure \ref{fig3}).   This transition curve would have been extremely difficult to obtain by numerically solving the NLD equation due to time feasibility and numerical constraints. 

As in the case of $\kappa=1$, we find very low frequency oscillations of the soliton position $q(t)$.  However, now we have a period of about $1400$ for 
$\kappa=2$ in the numerical simulations:  see figure \ref{fig4}. This has to be compared to the period $T = \pi/\omega(0) = 6.98$ of the fast oscillations. The CC theory agrees only qualitatively with the amplitude and period of the very low frequency oscillations, see bottom panel of figure 4. In the CC theory the fast oscillations are present but are not visible in the figure because their amplitude is too small. In the dissipative case when $\rho$ is nonzero, the charge $Q$  goes to zero and the soliton vanishes rather quickly even for small values of $\rho$.  For the particular value of $K$ and $r$ in figure 4 we display the fact that as we reduce $\kappa$ one goes from trapped behavior to unbound behavior. 
 
In conclusion, the CC theory is an excellent way of understanding the qualitative (and often quantitative)  behavior of the NLD equation solitons in various environments as a function of the parameters of the environment and the nonlinearity parameter $\kappa$.  In this paper we showed that it is quite useful in determining the transition curve from trapped to free behavior when there is a  spatially periodic parametric forcing term. 
 
 \appendix
 
 \section{Useful integrals} \label{appendix}
 
 Here, we calculate some useful integrals. Inserting the ansatz \eqref{psi1aXX}-\eqref{psi2aXX} in the following expression, and integrating over $x$ 
 \begin{eqnarray} \nonumber
 I_0(\omega)&=&- \int dx \left(\frac{i}{2}\right) [\bPsi \gamma^{1} \Psi_x 
 - \bPsi_x \gamma^{1} \Psi] %=\int dz \left( B' A-A'B  \right) 
 \\\label{I0}
 &&=\frac{\beta}{m+\omega} 
 \left[\frac{(\kappa+1)\beta^2}{g^2(m+\omega)} \right]^{1/\kappa} \, B\left(\frac{1}{2},1+\frac{1}{\kappa}\right)   
 \phantom{a}_2F_1\left(1+\frac{1}{\kappa},\frac{1}{2},\frac{3}{2}+\frac{1}{\kappa};\alpha^2\right),
 \end{eqnarray}
 where  $z=[x-q(t)] \cosh[\eta(t)]$, and $\alpha=\sqrt{\dfrac{m-\omega}{m+\omega}}$. 
 
 The effective potential \eqref{Ukappa} is defined as 
 \bq
 U=-  \int dx {\mathcal L}_3=- \dfrac{1}{\cosh[\eta(t)]} \int dz {\mathcal L}_3.     
 \eq
 By inserting the ansatz \eqref{psi1aXX}-\eqref{psi2aXX} in ${\mathcal L}_3$ we obtain 
 \begin{eqnarray}
 {\mathcal L}_3 &=&-\frac{1}{2} f \bPsi \Psi^\star - \frac{1}{2} f^\star \bPsi^\star \Psi=-\frac{r}{2} \cos \left(K \dfrac{z}{\cosh[\eta(t)]}+K\,q(t)\right) \left[A^2(z,t)+B^2(z,t)\right] \cos \left(2 \dfrac{p(t)}{\cosh[\eta(t)]} z-2 \phi
 (t)\right) \nonumber \\
 &=&-\frac{r}{2} \sum_{n=1}^2\, \left\{ \cos \left(2 \phi+ (-1)^{n}K\,q\right) \left[A^2(z,t)+B^2(z,t)\right] 
 \cos \left(\dfrac{2 p+(-1)^{n+1}K}{\cosh[\eta(t)]} z \right)\right\}.
 \end{eqnarray}
 Using the relation 
 \ba
 &&A^2 + B^2  = \left(\frac{  (\kappa+1)\beta ^2} {g^2\, \omega}\right)^{1/\kappa}  
 \left[  \frac{m}{\omega} \frac{ 1} {\left(\dfrac{m}{\omega}+ \cosh 2 \kappa \beta z\right)^{1/\kappa} } -
 \frac{\beta^2}{\omega^2} \frac{ 1} {\left(\dfrac{m}{\omega}+  \cosh 2 \kappa \beta z\right)^{ (\kappa+1)/\kappa }} \right],
 \ea
 we can write $U$ in terms of the function
 \bq \label{I-Integral}
 I_p[a,b,c,\nu] =  \int_0 ^\infty dz  ~ \frac{\cosh (i b z) }{ (a+  \cosh c z)^\nu}, 
 \eq
 with $a= m/\omega$, $c= 2 \kappa \beta$,  $\nu=1/\kappa$, and $b$ takes one of the two following values $b_{\pm} = (2p \pm K)  /\cosh \eta$. 
 This integral is found in   \cite{prudnikov:1986}.
 % Prudnikov  Integrals and Series Vol. 1(Eq. 9 on p. 357) 
 Note that  if we take a derivative with respect to $a$ we find
 \bq \label{trick}
 \frac {d I_p[a,b,c,\nu]}{da} = - \nu I_p[a,b,c,\nu+1].
 \eq
 Finally, we obtain 
 \ba
 &&I_p[a,b,c,\nu] = \frac{\Gamma \left(\nu -\frac{i b}{c}\right) \Gamma \left(\frac{i b}{c}+\nu
 	\right)}{\Gamma (\nu )} \times  \nonumber \\
 && \frac{\sqrt{\frac{\pi }{2}} \left(a+1\right)^{
 		\left(\frac{1}{2}-\nu \right)} \,
 	_2F_1\left(\frac{1}{2}-\frac{i b}{c},\frac{i b}{c}+\frac{1}{2};\nu
 	+\frac{1}{2};\frac{1-a}{2}\right)}{c \Gamma \left(\nu +\frac{1}{2}\right)} \,. 
 \ea
 %where 
 %\bq
 %a= m/\omega,  ~ b_\pm =(2p \pm K) /\cosh \eta, ~ c= 2 \kappa \beta, ~\nu=1/\kappa, \, \beta=\sqrt {m^2-\omega^2}.
 %\eq
 For $\kappa=1$, we have
 \bq
 I_p[a,b,c,1]= \frac{\pi  \text{csch}\left(\frac{\pi  b}{c}\right) \sin \left(\frac{b \cosh
 		^{-1}(a)}{c}\right)}{\sqrt{a^2-1} c} \,, 
 \eq
 so that
 \ba
 &&  \frac{m}{ \omega} I_p[a,b_{\pm} ,c,1] - \frac{\beta^2}{\omega^{2}}  I_p[a,b_{\pm} ,c,2]    \nonumber \\
 &&=\frac{\pi  (p \pm K/2) \text{sech}(\eta ) \text{csch}\left(\frac{\pi  (p \pm K/2)
 		\text{sech}(\eta )}{\beta}\right) \cos \left(\frac{(p \pm K/2)
 		\text{sech}(\eta ) \cosh ^{-1}\left(\frac{m}{\omega }\right)}{\beta}\right)}{2 \beta^2} \,. 
 \ea
 We find that  
 for $\kappa =1 $ our results simplify to the expression found in \cite{quintero2:2019}
 \ba
 && U=\frac{\pi  r \text{sech}^2(\eta )}{2 g^2 \omega } \nonumber \\
 &&  \times \left((2 p-K) \cos (K q +2 \phi ) \text{csch}\left(\frac{\pi  \text{sech}(\eta )
 	\left(p-\frac{K}{2}\right)}{\beta }\right) \cos \left(\frac{\text{sech}(\eta )
 	\left(p-\frac{K}{2}\right) \cosh ^{-1}\left(\frac{m}{\omega }\right)}{\beta
 }\right) \right. \nonumber \\
 &&+\left. (K+2 p) \cos (K q-2 \phi ) \text{csch}\left(\frac{\pi  \text{sech}(\eta )
 	\left(\frac{K}{2}+p\right)}{\beta }\right) \cos \left(\frac{\text{sech}(\eta )
 	\left(\frac{K}{2}+p\right) \cosh ^{-1}\left(\frac{m}{\omega }\right)}{\beta }\right) \right). \nonumber \\ 
 \ea
 At  $\kappa=1$ we have
 \bq
 Q=\frac{2 \beta}{g^2 \omega},
 \eq
 so the prefactor can also be written as 
 \bq
 \frac{\pi  r \text{sech}^2(\eta )}{2 g^2 \omega}  =\frac{r \pi Q \sech^2(\eta) }{4 \beta} \,. 
 \eq

 From our expression for $ I_0$, and $Q(\omega)$  we find that 
 \ba \label{rel1}
 &&\frac{d I_0}{d \omega} +Q(\omega)  =\frac{\sqrt{\pi } (\kappa +1) m \Gamma \left(1+\frac{1}{\kappa }\right) 
 	\left(\frac{(\kappa +1) (m-\omega )}{g^2}\right)^{\frac{1}{\kappa }}}{\kappa  \omega 
 	(m+\omega )^2 \sqrt{m^2-\omega ^2}}  N[\kappa,m,\omega], 
 \ea
 where
 \ba
 && N[\kappa,m,\omega] =
 (m+\omega )^2 \, _2\tilde{F}_1\left(-\frac{1}{2},1+\frac{1}{\kappa
 };\frac{3}{2}+\frac{1}{\kappa };\frac{m-\omega }{m+\omega }\right) \nonumber \\
 &&  -\omega  \left(2
 (m+\omega ) \, _2\tilde{F}_1\left(\frac{1}{2},1+\frac{1}{\kappa
 };\frac{3}{2}+\frac{1}{\kappa };\frac{m-\omega }{m+\omega }\right)  \right. \nonumber \\
 &&  \left.  +(m-\omega ) \,
 _2\tilde{F}_1\left(\frac{3}{2},2+\frac{1}{\kappa };\frac{5}{2}+\frac{1}{\kappa
 };\frac{m-\omega }{m+\omega } \right) \right)
 \equiv  0,
 \ea
 since this particular combination of hypergeometric functions sums to zero.  Here 
 $_2F_1[a,b; c;z]$ is the Gauss hypergeometric function and $_2\tilde{F}_1 [a,b;c;z]$ is the regularized hypergeometric function as defined by WolframMathworld.

 \begin{acknowledgments}
 	F.G.M. acknowledges financial support and hospitality of the University of Seville.  
 	N.R.Q. acknowledges the financial support from the Alexander von Humboldt Foundation 
 	and the hospitality of the Physikalisches Institut at the University of Bayreuth (Germany) and financial support from the Ministerio de Econom\'{\i}a y Competitividad of Spain through  FIS2017-89349-P. 
 	F.G.M. acknowledges financial support and hospitality of the Theoretical Division and Center for Nonlinear Studies at Los Alamos national Laboratory. F.C. would like to thank the Physics Department of Boston University for their hospitality while some of this work was performed.  This work was supported in part by the U.S. Department of Energy. 
 \end{acknowledgments}

\bibstyle{revtex}
 
\bibliography{dirac2a}

\begin{thebibliography}{41}%
\makeatletter
\providecommand \@ifxundefined [1]{%
 \@ifx{#1\undefined}
}%
\providecommand \@ifnum [1]{%
 \ifnum #1\expandafter \@firstoftwo
 \else \expandafter \@secondoftwo
 \fi
}%
\providecommand \@ifx [1]{%
 \ifx #1\expandafter \@firstoftwo
 \else \expandafter \@secondoftwo
 \fi
}%
\providecommand \natexlab [1]{#1}%
\providecommand \enquote  [1]{``#1''}%
\providecommand \bibnamefont  [1]{#1}%
\providecommand \bibfnamefont [1]{#1}%
\providecommand \citenamefont [1]{#1}%
\providecommand \href@noop [0]{\@secondoftwo}%
\providecommand \href [0]{\begingroup \@sanitize@url \@href}%
\providecommand \@href[1]{\@@startlink{#1}\@@href}%
\providecommand \@@href[1]{\endgroup#1\@@endlink}%
\providecommand \@sanitize@url [0]{\catcode `\\12\catcode `\$12\catcode
  `\&12\catcode `\#12\catcode `\^12\catcode `\_12\catcode `\%12\relax}%
\providecommand \@@startlink[1]{}%
\providecommand \@@endlink[0]{}%
\providecommand \url  [0]{\begingroup\@sanitize@url \@url }%
\providecommand \@url [1]{\endgroup\@href {#1}{\urlprefix }}%
\providecommand \urlprefix  [0]{URL }%
\providecommand \Eprint [0]{\href }%
\providecommand \doibase [0]{http://dx.doi.org/}%
\providecommand \selectlanguage [0]{\@gobble}%
\providecommand \bibinfo  [0]{\@secondoftwo}%
\providecommand \bibfield  [0]{\@secondoftwo}%
\providecommand \translation [1]{[#1]}%
\providecommand \BibitemOpen [0]{}%
\providecommand \bibitemStop [0]{}%
\providecommand \bibitemNoStop [0]{.\EOS\space}%
\providecommand \EOS [0]{\spacefactor3000\relax}%
\providecommand \BibitemShut  [1]{\csname bibitem#1\endcsname}%
\let\auto@bib@innerbib\@empty
%</preamble>
\bibitem [{\citenamefont {Fermi}(1934)}]{Fermi:1938}%
  \BibitemOpen
  \bibfield  {author} {\bibinfo {author} {\bibfnamefont {E.}~\bibnamefont
  {Fermi}},\ }\href@noop {} {\bibfield  {journal} {\bibinfo  {journal} {Il
  Nuovo Cimento}\ }\textbf {\bibinfo {volume} {11}},\ \bibinfo {pages} {1}
  (\bibinfo {year} {1934})}\BibitemShut {NoStop}%
\bibitem [{\citenamefont {R.P.Feynman}\ and\ \citenamefont
  {Gell-Mann}(1958)}]{Feynman:1958}%
  \BibitemOpen
  \bibfield  {author} {\bibinfo {author} {\bibnamefont {R.P.Feynman}}\ and\
  \bibinfo {author} {\bibfnamefont {M.}~\bibnamefont {Gell-Mann}},\ }\href@noop
  {} {\bibfield  {journal} {\bibinfo  {journal} {Physical Review}\ }\textbf
  {\bibinfo {volume} {109}},\ \bibinfo {pages} {193} (\bibinfo {year}
  {1958})}\BibitemShut {NoStop}%
\bibitem [{\citenamefont {Ivanenko}(1938)}]{Ivanenko:1938}%
  \BibitemOpen
  \bibfield  {author} {\bibinfo {author} {\bibfnamefont {D.~D.}\ \bibnamefont
  {Ivanenko}},\ }\href@noop {} {\bibfield  {journal} {\bibinfo  {journal} {Zh.
  Eksp. Teor. Fiz}\ }\textbf {\bibinfo {volume} {8}},\ \bibinfo {pages} {260}
  (\bibinfo {year} {1938})}\BibitemShut {NoStop}%
\bibitem [{\citenamefont {{R. Finkelstein, R. Lelevier, and M.
  Ruderman}}(1951)}]{Finkelstein:1951}%
  \BibitemOpen
  \bibfield  {author} {\bibinfo {author} {\bibnamefont {{R. Finkelstein, R.
  Lelevier, and M. Ruderman}}},\ }\href@noop {} {\bibfield  {journal} {\bibinfo
   {journal} {Phys. Rev. D}\ }\textbf {\bibinfo {volume} {83}},\ \bibinfo
  {pages} {326} (\bibinfo {year} {1951})}\BibitemShut {NoStop}%
\bibitem [{\citenamefont {Finkelstein}\ \emph {et~al.}(1956)\citenamefont
  {Finkelstein}, \citenamefont {Fronsdal},\ and\ \citenamefont
  {Kaus}}]{finkelstein:1956}%
  \BibitemOpen
  \bibfield  {author} {\bibinfo {author} {\bibfnamefont {R.}~\bibnamefont
  {Finkelstein}}, \bibinfo {author} {\bibfnamefont {C.}~\bibnamefont
  {Fronsdal}}, \ and\ \bibinfo {author} {\bibfnamefont {P.}~\bibnamefont
  {Kaus}},\ }\href@noop {} {\bibfield  {journal} {\bibinfo  {journal} {Phys.
  Rev.}\ }\textbf {\bibinfo {volume} {103}},\ \bibinfo {pages} {1571} (\bibinfo
  {year} {1956})}\BibitemShut {NoStop}%
\bibitem [{\citenamefont {Heisenberg}(1957)}]{heisenberg:1957}%
  \BibitemOpen
  \bibfield  {author} {\bibinfo {author} {\bibfnamefont {W.}~\bibnamefont
  {Heisenberg}},\ }\href@noop {} {\bibfield  {journal} {\bibinfo  {journal}
  {Rev. Mod. Phys.}\ }\textbf {\bibinfo {volume} {29}},\ \bibinfo {pages} {269}
  (\bibinfo {year} {1957})}\BibitemShut {NoStop}%
\bibitem [{\citenamefont {Barashenkov}\ \emph {et~al.}(1998)\citenamefont
  {Barashenkov}, \citenamefont {Pelinovsky},\ and\ \citenamefont
  {Zemlyanaya}}]{barashenkov:1998}%
  \BibitemOpen
  \bibfield  {author} {\bibinfo {author} {\bibfnamefont {I.~V.}\ \bibnamefont
  {Barashenkov}}, \bibinfo {author} {\bibfnamefont {D.~E.}\ \bibnamefont
  {Pelinovsky}}, \ and\ \bibinfo {author} {\bibfnamefont {E.~V.}\ \bibnamefont
  {Zemlyanaya}},\ }\href@noop {} {\bibfield  {journal} {\bibinfo  {journal}
  {Phys. Rev. Lett.}\ }\textbf {\bibinfo {volume} {80}},\ \bibinfo {pages}
  {5117} (\bibinfo {year} {1998})}\BibitemShut {NoStop}%
\bibitem [{\citenamefont {Longhi}(2010)}]{longhi:2010}%
  \BibitemOpen
  \bibfield  {author} {\bibinfo {author} {\bibfnamefont {S.}~\bibnamefont
  {Longhi}},\ }\href@noop {} {\bibfield  {journal} {\bibinfo  {journal} {Opt.
  Lett.}\ }\textbf {\bibinfo {volume} {35}},\ \bibinfo {pages} {235} (\bibinfo
  {year} {2010})}\BibitemShut {NoStop}%
\bibitem [{\citenamefont {Dreisow}\ \emph {et~al.}(2010)\citenamefont
  {Dreisow}, \citenamefont {Heinrich}, \citenamefont {Keil}, \citenamefont
  {T\"unnermann}, \citenamefont {Nolte}, \citenamefont {Longhi},\ and\
  \citenamefont {Szameit}}]{dreisow:2010}%
  \BibitemOpen
  \bibfield  {author} {\bibinfo {author} {\bibfnamefont {F.}~\bibnamefont
  {Dreisow}}, \bibinfo {author} {\bibfnamefont {M.}~\bibnamefont {Heinrich}},
  \bibinfo {author} {\bibfnamefont {R.}~\bibnamefont {Keil}}, \bibinfo {author}
  {\bibfnamefont {A.}~\bibnamefont {T\"unnermann}}, \bibinfo {author}
  {\bibfnamefont {S.}~\bibnamefont {Nolte}}, \bibinfo {author} {\bibfnamefont
  {S.}~\bibnamefont {Longhi}}, \ and\ \bibinfo {author} {\bibfnamefont
  {A.}~\bibnamefont {Szameit}},\ }\href {\doibase
  10.1103/PhysRevLett.105.143902} {\bibfield  {journal} {\bibinfo  {journal}
  {Phys. Rev. Lett.}\ }\textbf {\bibinfo {volume} {105}},\ \bibinfo {pages}
  {143902} (\bibinfo {year} {2010})}\BibitemShut {NoStop}%
\bibitem [{\citenamefont {Truong~X.}\ \emph {et~al.}(2014)\citenamefont
  {Truong~X.}, \citenamefont {Stefano},\ and\ \citenamefont
  {Fabio}}]{tran:2014}%
  \BibitemOpen
  \bibfield  {author} {\bibinfo {author} {\bibfnamefont {T.}~\bibnamefont
  {Truong~X.}}, \bibinfo {author} {\bibfnamefont {L.}~\bibnamefont {Stefano}},
  \ and\ \bibinfo {author} {\bibfnamefont {B.}~\bibnamefont {Fabio}},\
  }\href@noop {} {\bibfield  {journal} {\bibinfo  {journal} {Annals of
  Physics}\ }\textbf {\bibinfo {volume} {340}},\ \bibinfo {pages} {179}
  (\bibinfo {year} {2014})}\BibitemShut {NoStop}%
\bibitem [{\citenamefont {Haddad}\ and\ \citenamefont
  {Carr}(2009)}]{haddad:2009}%
  \BibitemOpen
  \bibfield  {author} {\bibinfo {author} {\bibfnamefont {L.}~\bibnamefont
  {Haddad}}\ and\ \bibinfo {author} {\bibfnamefont {L.}~\bibnamefont {Carr}},\
  }\href@noop {} {\bibfield  {journal} {\bibinfo  {journal} {Phys. D: Nonlinear
  Phenomena}\ }\textbf {\bibinfo {volume} {238}},\ \bibinfo {pages} {1413}
  (\bibinfo {year} {2009})}\BibitemShut {NoStop}%
\bibitem [{\citenamefont {Ra\~nada}\ and\ \citenamefont
  {Ra\~nada}(1984)}]{ranada:1984}%
  \BibitemOpen
  \bibfield  {author} {\bibinfo {author} {\bibfnamefont {A.~F.}\ \bibnamefont
  {Ra\~nada}}\ and\ \bibinfo {author} {\bibfnamefont {M.~F.}\ \bibnamefont
  {Ra\~nada}},\ }\href {\doibase 10.1103/PhysRevD.29.985} {\bibfield  {journal}
  {\bibinfo  {journal} {Phys. Rev. D}\ }\textbf {\bibinfo {volume} {29}},\
  \bibinfo {pages} {985} (\bibinfo {year} {1984})}\BibitemShut {NoStop}%
\bibitem [{\citenamefont {Weyl}(1950)}]{weyl:1950}%
  \BibitemOpen
  \bibfield  {author} {\bibinfo {author} {\bibfnamefont {H.}~\bibnamefont
  {Weyl}},\ }\href@noop {} {\bibfield  {journal} {\bibinfo  {journal} {Phys.
  Rev.}\ }\textbf {\bibinfo {volume} {77}},\ \bibinfo {pages} {699} (\bibinfo
  {year} {1950})}\BibitemShut {NoStop}%
\bibitem [{\citenamefont {Lee}\ \emph {et~al.}(1975)\citenamefont {Lee},
  \citenamefont {Kuo},\ and\ \citenamefont {Gavrielides}}]{lee:1975}%
  \BibitemOpen
  \bibfield  {author} {\bibinfo {author} {\bibfnamefont {S.~Y.}\ \bibnamefont
  {Lee}}, \bibinfo {author} {\bibfnamefont {T.~K.}\ \bibnamefont {Kuo}}, \ and\
  \bibinfo {author} {\bibfnamefont {A.}~\bibnamefont {Gavrielides}},\
  }\href@noop {} {\bibfield  {journal} {\bibinfo  {journal} {Phys. Rev. D}\
  }\textbf {\bibinfo {volume} {12}},\ \bibinfo {pages} {2249} (\bibinfo {year}
  {1975})}\BibitemShut {NoStop}%
\bibitem [{\citenamefont {Chang}\ \emph {et~al.}(1975)\citenamefont {Chang},
  \citenamefont {Ellis},\ and\ \citenamefont {Lee}}]{chang:1975}%
  \BibitemOpen
  \bibfield  {author} {\bibinfo {author} {\bibfnamefont {S.-J.}\ \bibnamefont
  {Chang}}, \bibinfo {author} {\bibfnamefont {S.~D.}\ \bibnamefont {Ellis}}, \
  and\ \bibinfo {author} {\bibfnamefont {B.~W.}\ \bibnamefont {Lee}},\ }\href
  {\doibase 10.1103/PhysRevD.11.3572} {\bibfield  {journal} {\bibinfo
  {journal} {Phys. Rev. D}\ }\textbf {\bibinfo {volume} {11}},\ \bibinfo
  {pages} {3572} (\bibinfo {year} {1975})}\BibitemShut {NoStop}%
\bibitem [{\citenamefont {Mathieu}(1985{\natexlab{a}})}]{mathieuprd:1985}%
  \BibitemOpen
  \bibfield  {author} {\bibinfo {author} {\bibfnamefont {P.}~\bibnamefont
  {Mathieu}},\ }\href@noop {} {\bibfield  {journal} {\bibinfo  {journal} {Phys.
  Rev. D}\ }\textbf {\bibinfo {volume} {32}},\ \bibinfo {pages} {3288}
  (\bibinfo {year} {1985}{\natexlab{a}})}\BibitemShut {NoStop}%
\bibitem [{\citenamefont {Stubbe}(1986)}]{stubbe:1986}%
  \BibitemOpen
  \bibfield  {author} {\bibinfo {author} {\bibfnamefont {J.}~\bibnamefont
  {Stubbe}},\ }\href@noop {} {\bibfield  {journal} {\bibinfo  {journal} {J.
  Math. Phys.}\ }\textbf {\bibinfo {volume} {27}},\ \bibinfo {pages} {2561}
  (\bibinfo {year} {1986})}\BibitemShut {NoStop}%
\bibitem [{\citenamefont {Cooper}\ \emph {et~al.}(2010)\citenamefont {Cooper},
  \citenamefont {Khare}, \citenamefont {Mihaila},\ and\ \citenamefont
  {Saxena}}]{cooper:2010}%
  \BibitemOpen
  \bibfield  {author} {\bibinfo {author} {\bibfnamefont {F.}~\bibnamefont
  {Cooper}}, \bibinfo {author} {\bibfnamefont {A.}~\bibnamefont {Khare}},
  \bibinfo {author} {\bibfnamefont {B.}~\bibnamefont {Mihaila}}, \ and\
  \bibinfo {author} {\bibfnamefont {A.}~\bibnamefont {Saxena}},\ }\href@noop {}
  {\bibfield  {journal} {\bibinfo  {journal} {Phys. Rev. E}\ }\textbf {\bibinfo
  {volume} {82}},\ \bibinfo {pages} {036604} (\bibinfo {year}
  {2010})}\BibitemShut {NoStop}%
\bibitem [{\citenamefont {{J. Xu, and S. Shao, and H. Tang}}(2013)}]{xu:2013}%
  \BibitemOpen
  \bibfield  {author} {\bibinfo {author} {\bibnamefont {{J. Xu, and S. Shao,
  and H. Tang}}},\ }\href {\doibase https://doi.org/10.1016/j.jcp.2013.03.031}
  {\bibfield  {journal} {\bibinfo  {journal} {J. Comput. Phys.}\ }\textbf
  {\bibinfo {volume} {245}},\ \bibinfo {pages} {131} (\bibinfo {year}
  {2013})}\BibitemShut {NoStop}%
\bibitem [{\citenamefont {Mathieu}(1985{\natexlab{b}})}]{mathieu:1985}%
  \BibitemOpen
  \bibfield  {author} {\bibinfo {author} {\bibfnamefont {P.}~\bibnamefont
  {Mathieu}},\ }\href {http://stacks.iop.org/0305-4470/18/i=16/a=012}
  {\bibfield  {journal} {\bibinfo  {journal} {Journal of Physics A:
  Mathematical and General}\ }\textbf {\bibinfo {volume} {18}},\ \bibinfo
  {pages} {L1061} (\bibinfo {year} {1985}{\natexlab{b}})}\BibitemShut {NoStop}%
\bibitem [{\citenamefont {Alvarez}\ and\ \citenamefont
  {Carreras}(1981)}]{alvarez:1981}%
  \BibitemOpen
  \bibfield  {author} {\bibinfo {author} {\bibfnamefont {A.}~\bibnamefont
  {Alvarez}}\ and\ \bibinfo {author} {\bibfnamefont {B.}~\bibnamefont
  {Carreras}},\ }\href@noop {} {\bibfield  {journal} {\bibinfo  {journal}
  {Phys. Lett. A}\ }\textbf {\bibinfo {volume} {86}},\ \bibinfo {pages} {327}
  (\bibinfo {year} {1981})}\BibitemShut {NoStop}%
\bibitem [{\citenamefont {Shao}\ and\ \citenamefont {Tang}(2005)}]{shao:2005}%
  \BibitemOpen
  \bibfield  {author} {\bibinfo {author} {\bibfnamefont {S.}~\bibnamefont
  {Shao}}\ and\ \bibinfo {author} {\bibfnamefont {H.}~\bibnamefont {Tang}},\
  }\href@noop {} {\bibfield  {journal} {\bibinfo  {journal} {Phys. Lett. A}\
  }\textbf {\bibinfo {volume} {345}},\ \bibinfo {pages} {119} (\bibinfo {year}
  {2005})}\BibitemShut {NoStop}%
\bibitem [{\citenamefont {Shao}\ and\ \citenamefont {Tang}(2006)}]{shao:2006}%
  \BibitemOpen
  \bibfield  {author} {\bibinfo {author} {\bibfnamefont {S.~H.}\ \bibnamefont
  {Shao}}\ and\ \bibinfo {author} {\bibfnamefont {H.}~\bibnamefont {Tang}},\
  }\href@noop {} {\bibfield  {journal} {\bibinfo  {journal} {Discrete Cont.
  Dyn. Syst. B}\ }\textbf {\bibinfo {volume} {6}},\ \bibinfo {pages} {623}
  (\bibinfo {year} {2006})}\BibitemShut {NoStop}%
\bibitem [{\citenamefont {{S. H. Shao and H. Z. Tang}}(2008)}]{shao:2008}%
  \BibitemOpen
  \bibfield  {author} {\bibinfo {author} {\bibnamefont {{S. H. Shao and H. Z.
  Tang}}},\ }\href@noop {} {\bibfield  {journal} {\bibinfo  {journal} {Commun.
  Comput. Phys.}\ }\textbf {\bibinfo {volume} {3}},\ \bibinfo {pages} {950}
  (\bibinfo {year} {2008})}\BibitemShut {NoStop}%
\bibitem [{\citenamefont {Scharf}\ \emph {et~al.}(1992)\citenamefont {Scharf},
  \citenamefont {Kivshar}, \citenamefont {S\'{a}nchez},\ and\ \citenamefont
  {Bishop}}]{scharf:1992}%
  \BibitemOpen
  \bibfield  {author} {\bibinfo {author} {\bibfnamefont {R.}~\bibnamefont
  {Scharf}}, \bibinfo {author} {\bibfnamefont {Y.~S.}\ \bibnamefont {Kivshar}},
  \bibinfo {author} {\bibfnamefont {A.}~\bibnamefont {S\'{a}nchez}}, \ and\
  \bibinfo {author} {\bibfnamefont {A.~R.}\ \bibnamefont {Bishop}},\ }\href
  {\doibase 10.1103/PhysRevA.45.R5369} {\bibfield  {journal} {\bibinfo
  {journal} {Phys. Rev. A}\ }\textbf {\bibinfo {volume} {45}},\ \bibinfo
  {pages} {R5369} (\bibinfo {year} {1992})}\BibitemShut {NoStop}%
\bibitem [{\citenamefont {{Angel, S\'{a}nchez and Rainer, Scharf and Alan R.,
  Bishop and Luis, V\'azquez}}(1992)}]{sanchez:1992}%
  \BibitemOpen
  \bibfield  {author} {\bibinfo {author} {\bibnamefont {{Angel, S\'{a}nchez and
  Rainer, Scharf and Alan R., Bishop and Luis, V\'azquez}}},\ }\href {\doibase
  10.1103/PhysRevA.45.6031} {\bibfield  {journal} {\bibinfo  {journal} {Phys.
  Rev. A}\ }\textbf {\bibinfo {volume} {45}},\ \bibinfo {pages} {6031}
  (\bibinfo {year} {1992})}\BibitemShut {NoStop}%
\bibitem [{\citenamefont {Cuenda}\ and\ \citenamefont
  {S\'{a}nchez}(2005)}]{cuenda:2005}%
  \BibitemOpen
  \bibfield  {author} {\bibinfo {author} {\bibfnamefont {S.}~\bibnamefont
  {Cuenda}}\ and\ \bibinfo {author} {\bibfnamefont {A.}~\bibnamefont
  {S\'{a}nchez}},\ }\href@noop {} {\bibfield  {journal} {\bibinfo  {journal}
  {Chaos}\ }\textbf {\bibinfo {volume} {15}},\ \bibinfo {pages} {023502}
  (\bibinfo {year} {2005})}\BibitemShut {NoStop}%
\bibitem [{\citenamefont {S\'{a}nchez}\ \emph {et~al.}(1994)\citenamefont
  {S\'{a}nchez}, \citenamefont {Bishop},\ and\ \citenamefont
  {Dom\'{i}nguez-Adame}}]{sanchez:1994}%
  \BibitemOpen
  \bibfield  {author} {\bibinfo {author} {\bibfnamefont {A.}~\bibnamefont
  {S\'{a}nchez}}, \bibinfo {author} {\bibfnamefont {A.~R.}\ \bibnamefont
  {Bishop}}, \ and\ \bibinfo {author} {\bibfnamefont {F.}~\bibnamefont
  {Dom\'{i}nguez-Adame}},\ }\href {\doibase 10.1103/PhysRevE.49.4603}
  {\bibfield  {journal} {\bibinfo  {journal} {Phys. Rev. E}\ }\textbf {\bibinfo
  {volume} {{49}}},\ \bibinfo {pages} {{4603}} (\bibinfo {year}
  {{1994}})}\BibitemShut {NoStop}%
\bibitem [{\citenamefont {Scharf}\ and\ \citenamefont
  {Bishop}(1993)}]{scharf:1993}%
  \BibitemOpen
  \bibfield  {author} {\bibinfo {author} {\bibfnamefont {R.}~\bibnamefont
  {Scharf}}\ and\ \bibinfo {author} {\bibfnamefont {A.~R.}\ \bibnamefont
  {Bishop}},\ }\href {\doibase 10.1103/PhysRevE.47.1375} {\bibfield  {journal}
  {\bibinfo  {journal} {Phys. Rev. E}\ }\textbf {\bibinfo {volume} {47}},\
  \bibinfo {pages} {1375} (\bibinfo {year} {1993})}\BibitemShut {NoStop}%
\bibitem [{\citenamefont {{A. Bondeson, M. Lisak and D.
  Anderson}}(1979)}]{bondeson:1979}%
  \BibitemOpen
  \bibfield  {author} {\bibinfo {author} {\bibnamefont {{A. Bondeson, M. Lisak
  and D. Anderson}}},\ }\href@noop {} {\bibfield  {journal} {\bibinfo
  {journal} {Physica Scripta}\ }\textbf {\bibinfo {volume} {20}},\ \bibinfo
  {pages} {479} (\bibinfo {year} {1979})}\BibitemShut {NoStop}%
\bibitem [{\citenamefont {Kishvar}(1990)}]{kishvar:1990}%
  \BibitemOpen
  \bibfield  {author} {\bibinfo {author} {\bibfnamefont {Y.}~\bibnamefont
  {Kishvar}},\ }\href@noop {} {\bibfield  {journal} {\bibinfo  {journal} {J.
  Opt. Soc. Am B}\ }\textbf {\bibinfo {volume} {7}},\ \bibinfo {pages} {2204}
  (\bibinfo {year} {1990})}\BibitemShut {NoStop}%
\bibitem [{\citenamefont {{F.Cooper, C. Lucheroni and H.
  Shepard}}(1992)}]{cooper:1992}%
  \BibitemOpen
  \bibfield  {author} {\bibinfo {author} {\bibnamefont {{F.Cooper, C. Lucheroni
  and H. Shepard}}},\ }\href@noop {} {\bibfield  {journal} {\bibinfo  {journal}
  {Phys. Lett. A}\ }\textbf {\bibinfo {volume} {170}},\ \bibinfo {pages} {184}
  (\bibinfo {year} {1992})}\BibitemShut {NoStop}%
\bibitem [{\citenamefont {Cooper}\ \emph
  {et~al.}(1993{\natexlab{a}})\citenamefont {Cooper}, \citenamefont
  {Lucheroni}, \citenamefont {Shepard},\ and\ \citenamefont
  {Sodano}}]{cooper:1993a}%
  \BibitemOpen
  \bibfield  {author} {\bibinfo {author} {\bibfnamefont {F.}~\bibnamefont
  {Cooper}}, \bibinfo {author} {\bibfnamefont {C.}~\bibnamefont {Lucheroni}},
  \bibinfo {author} {\bibfnamefont {H.}~\bibnamefont {Shepard}}, \ and\
  \bibinfo {author} {\bibfnamefont {P.}~\bibnamefont {Sodano}},\ }\href@noop {}
  {\bibfield  {journal} {\bibinfo  {journal} {Physica D}\ }\textbf {\bibinfo
  {volume} {68}},\ \bibinfo {pages} {344} (\bibinfo {year}
  {1993}{\natexlab{a}})}\BibitemShut {NoStop}%
\bibitem [{\citenamefont {Cooper}\ \emph
  {et~al.}(1993{\natexlab{b}})\citenamefont {Cooper}, \citenamefont
  {Lucheroni}, \citenamefont {Shepard},\ and\ \citenamefont
  {Sodano}}]{cooper:1993b}%
  \BibitemOpen
  \bibfield  {author} {\bibinfo {author} {\bibfnamefont {F.}~\bibnamefont
  {Cooper}}, \bibinfo {author} {\bibfnamefont {C.}~\bibnamefont {Lucheroni}},
  \bibinfo {author} {\bibfnamefont {H.}~\bibnamefont {Shepard}}, \ and\
  \bibinfo {author} {\bibfnamefont {P.}~\bibnamefont {Sodano}},\ }\href@noop {}
  {\bibfield  {journal} {\bibinfo  {journal} {Phys. Lett. A}\ ,\ \bibinfo
  {pages} {33}} (\bibinfo {year} {1993}{\natexlab{b}})}\BibitemShut {NoStop}%
\bibitem [{\citenamefont {Cooper}\ \emph
  {et~al.}(1993{\natexlab{c}})\citenamefont {Cooper}, \citenamefont {Shepard},\
  and\ \citenamefont {Sodano}}]{cooper:1993c}%
  \BibitemOpen
  \bibfield  {author} {\bibinfo {author} {\bibfnamefont {F.}~\bibnamefont
  {Cooper}}, \bibinfo {author} {\bibfnamefont {H.}~\bibnamefont {Shepard}}, \
  and\ \bibinfo {author} {\bibfnamefont {P.}~\bibnamefont {Sodano}},\
  }\href@noop {} {\bibfield  {journal} {\bibinfo  {journal} {Phys. Rev. E}\
  }\textbf {\bibinfo {volume} {48}},\ \bibinfo {pages} {4027} (\bibinfo {year}
  {1993}{\natexlab{c}})}\BibitemShut {NoStop}%
\bibitem [{\citenamefont {Cooper}\ and\ \citenamefont
  {Shepard}(1994)}]{cooper:1994}%
  \BibitemOpen
  \bibfield  {author} {\bibinfo {author} {\bibfnamefont {F.}~\bibnamefont
  {Cooper}}\ and\ \bibinfo {author} {\bibfnamefont {H.}~\bibnamefont
  {Shepard}},\ }\href@noop {} {\bibfield  {journal} {\bibinfo  {journal} {Phys.
  Lett. A}\ }\textbf {\bibinfo {volume} {194}},\ \bibinfo {pages} {246}
  (\bibinfo {year} {1994})}\BibitemShut {NoStop}%
\bibitem [{\citenamefont {Cooper}\ \emph {et~al.}(2017)\citenamefont {Cooper},
  \citenamefont {Dawson}, \citenamefont {Mertens}, \citenamefont {Arevalo},
  \citenamefont {Quintero}, \citenamefont {Mihaila}, \citenamefont {Khare},\
  and\ \citenamefont {Saxena}}]{cooper:2017}%
  \BibitemOpen
  \bibfield  {author} {\bibinfo {author} {\bibfnamefont {F.}~\bibnamefont
  {Cooper}}, \bibinfo {author} {\bibfnamefont {J.}~\bibnamefont {Dawson}},
  \bibinfo {author} {\bibfnamefont {F.}~\bibnamefont {Mertens}}, \bibinfo
  {author} {\bibfnamefont {E.}~\bibnamefont {Arevalo}}, \bibinfo {author}
  {\bibfnamefont {N.}~\bibnamefont {Quintero}}, \bibinfo {author}
  {\bibfnamefont {B.}~\bibnamefont {Mihaila}}, \bibinfo {author} {\bibfnamefont
  {A.}~\bibnamefont {Khare}}, \ and\ \bibinfo {author} {\bibfnamefont
  {A.}~\bibnamefont {Saxena}},\ }\href@noop {} {\bibfield  {journal} {\bibinfo
  {journal} {Journal of Physics A: Mathematical and Theoretical}\ }\textbf
  {\bibinfo {volume} {50}},\ \bibinfo {pages} {485205} (\bibinfo {year}
  {2017})}\BibitemShut {NoStop}%
\bibitem [{\citenamefont {Quintero}\ \emph
  {et~al.}(2019{\natexlab{a}})\citenamefont {Quintero}, \citenamefont
  {Sanchez-Rey}, \citenamefont {Cooper},\ and\ \citenamefont
  {Mertens}}]{quintero2:2019}%
  \BibitemOpen
  \bibfield  {author} {\bibinfo {author} {\bibfnamefont {N.~R.}\ \bibnamefont
  {Quintero}}, \bibinfo {author} {\bibfnamefont {B.}~\bibnamefont
  {Sanchez-Rey}}, \bibinfo {author} {\bibfnamefont {F.}~\bibnamefont {Cooper}},
  \ and\ \bibinfo {author} {\bibfnamefont {F.~G.}\ \bibnamefont {Mertens}},\
  }\href@noop {} {\bibfield  {journal} {\bibinfo  {journal} {J. Phys. A: Math.
  Gen.}\ }\textbf {\bibinfo {volume} {52}},\ \bibinfo {pages} {285201}
  (\bibinfo {year} {2019}{\natexlab{a}})}\BibitemShut {NoStop}%
\bibitem [{\citenamefont {Quintero}\ \emph
  {et~al.}(2019{\natexlab{b}})\citenamefont {Quintero}, \citenamefont {Shao},
  \citenamefont {Alvarez-Nodarse},\ and\ \citenamefont
  {Mertens}}]{quintero:2019}%
  \BibitemOpen
  \bibfield  {author} {\bibinfo {author} {\bibfnamefont {N.~R.}\ \bibnamefont
  {Quintero}}, \bibinfo {author} {\bibfnamefont {S.}~\bibnamefont {Shao}},
  \bibinfo {author} {\bibfnamefont {R.}~\bibnamefont {Alvarez-Nodarse}}, \ and\
  \bibinfo {author} {\bibfnamefont {F.~G.}\ \bibnamefont {Mertens}},\
  }\href@noop {} {\bibfield  {journal} {\bibinfo  {journal} {J. Phys. A: Math.
  Gen.}\ }\textbf {\bibinfo {volume} {52}},\ \bibinfo {pages} {155401}
  (\bibinfo {year} {2019}{\natexlab{b}})}\BibitemShut {NoStop}%
\bibitem [{\citenamefont {Rayleigh}(1877)}]{Rayleigh}%
  \BibitemOpen
  \bibfield  {author} {\bibinfo {author} {\bibfnamefont {L.}~\bibnamefont
  {Rayleigh}},\ }\href@noop {} {\emph {\bibinfo {title} {Theory of Sound}}}\
  (\bibinfo  {publisher} {MacMillan and Company},\ \bibinfo {address}
  {London},\ \bibinfo {year} {1877})\BibitemShut {NoStop}%
\bibitem [{\citenamefont {{A.P. Prudnikov, Y.A. Brychkov and O.I.
  Marichev}}(1986)}]{prudnikov:1986}%
  \BibitemOpen
  \bibfield  {author} {\bibinfo {author} {\bibnamefont {{A.P. Prudnikov, Y.A.
  Brychkov and O.I. Marichev}}},\ }\href@noop {} {\emph {\bibinfo {title}
  {Integrals and Series Vol. I, Elementary Functions}}}\ (\bibinfo  {publisher}
  {Gordon and Breach Scientific Publishers},\ \bibinfo {year}
  {1986})\BibitemShut {NoStop}%
\end{thebibliography}%

\end{document}